\documentclass[iop,preprint,10pt]{emulateapj}
\usepackage[]{natbib}
\usepackage{graphicx}
\usepackage{subfigure}
\usepackage[dvipsnames]{color}
\usepackage{float}
\usepackage{booktabs}
\usepackage{amsmath}
\usepackage{fancyref}
\definecolor{orange}{RGB}{255,69,0}
\definecolor{green}{RGB}{0,255,0}
\definecolor{darkred}{RGB}{139,0,0}
\usepackage{amssymb}
\usepackage{lineno}
\usepackage[colorlinks=true,
            linkcolor=orange,
            urlcolor=magenta,
            citecolor=blue]{hyperref}
\usepackage{nameref}
\usepackage{array,multirow,makecell}
\usepackage{tabularx}

\usepackage{morefloats}
\usepackage{gensymb}
\begin{document}
\title
{Two-zone emission modeling of PKS 1510-089 during the high state of 2015 }

\author{Raj Prince$^1$, Nayantara Gupta$^1$, Krzysztof Nalewajko$^2$}
\affil{$^1$Raman Research Institute, C.V. Raman Avenue, Sadashivanagar, Bangalore 560080, India \\
$^2$Nicolaus Copernicus Astronomical Center, Polish Academy of Sciences, Bartycka 18, 00-716 Warsaw, Poland}

\email{rajprince@rri.res.in}
\begin{abstract}
PKS 1510-089 is one of the most variable blazars in the third \textit{Fermi}-LAT source catalog. During 2015, this source has shown
four flares identified as flare A, B, C, and D in between three quiescent states Q1, Q2, and Q3.  The multiwavelength data from \textit{Fermi}-LAT, 
Swift-XRT/UVOT, OVRO, and SMA Observatory are used in our work to model these states. Different flux doubling times have been observed 
in different energy bands which indicate there could be multiple emission zones. The flux doubling time from the gamma-ray and
X-ray light curve are found to be 10.6 hr, 2.5 days, and the average flux doubling time in the optical/UV band is 1 day.
It is possible that the gamma-ray and optical/UV emission are produced in the same region whereas X-ray emission is coming from a 
different region along the jet axis. We have also estimated the discrete correlations functions (DCFs) among the light curves of different energy bands
to infer about their emission regions. However, our DCF analysis does not show significant correlation in different energy bands though
it shows peaks in some cases at small time lags. 
We perform a two-zone multi-wavelength time-dependent modeling  with one emission zone located near
the outer edge of the broad line region (BLR) and another further away in the dusty/molecular torus (DT/MT) region to study this high state.

\end{abstract}

\keywords{galaxies: active; gamma-rays: galaxies; individuals: PKS 1510-089}

\section{Introduction}

Being one of the most variable flat spectrum radio quasars (FSRQs) in the third \textit{Fermi}-LAT source catalog (3FGL) PKS 1510-089 has been well
observed during its high states 
in the past. It is located at a redshift of z = 0.361 \citep{Tanner et al. (1996)} with black hole mass 5.4$\times 10^8$ M$_{\sun}$ and accretion 
rate approximately 0.5 M$_{\sun}$/yr (\citealt{Abdo et al. (2010)}). 
A long term analysis of the light curve of PKS 1510-089 with the eight year \textit{Fermi}-LAT data has been done by \citet{Prince et al. (2017)} earlier. They 
have observed five major flares during 2008-2016, and their temporal and spectral features have been studied in detail.
During its high activity period between September 2008 and June 2009 its gamma-ray emission showed a weak correlation with the UV, strong correlation
with the optical and no correlation with the X-ray emission \citep{Abdo et al. (2010)}. PKS 1510-089 was also studied by \citet{Nalewajko (2013)}, where
he used the first four years of \textit{Fermi}-LAT data and observed 14 flares with a minimum and maximum flux of 7.4 and 26.6 ($\times$10$^{-6}$)
ph cm$^{-2}$ s$^{-1}$, respectively. 
Detection of high energy gamma-rays up to 300-400 GeV has been
reported by the H.E.S.S. collaboration (\citealt{Abramowski et al. (2013)}) during March-April 2009 and by the MAGIC collaboration 
(\citealt{Aleksic et al. (2014)}) between February 3 and April 3, 2012.

In the second half of 2011 this source was active in several energy bands and the optical, gamma-ray and radio flares were detected. The gamma-ray 
variability 
down to 20 minutes indicated the highly variable nature of this source. \citealt{Aleksic et al. (2014)} did a detailed multiwavelength modeling for
the period January-April 2012 covering radio to very high energy gamma-rays.
They explained the multiwavelength emission as the result of turbulent plasma flowing  at a relativistic speed  down the jet and crossing a standing 
conical  shock.
In modeling the spectral energy distributions (SEDs) from PKS 1510-089 it is most commonly thought that the low energy (radio, optical) emission is 
from synchrotron radiation of relativistic electrons and high energy emission (X-ray, gamma-ray) is from external Compton (EC) scattering of 
the seed photons in the broad line region (BLR) and dusty torus region (\citealt{Kataoka et al. (2008)}, \citealt {Abdo et al. (2010)},
\citealt{Brown (2013)}, \citealt{Barnacka et al. (2014)}). The gamma-ray emission region could also be located at a radio knot, far away from 
the black hole as suggested by \citet{Marscher et al. (2010)}. They modeled the eight major gamma-ray flares of PKS 1510-089  that happened in 
2009. During optical and gamma-ray flare a bright radio knot travelled through the core/stationary feature at 43 GHz seen by VLBA (Very Long 
Baseline  Array) images. The knot continued to propagate down the length of the jet at an apparent speed of 22c. A strong emission in gamma-ray 
energy accompanied by a month long emission in X-ray/radio emission which gradually intensified, represented the complex nature of  the flares. 

The hadronic scenario of high energy photon emission (X-ray, gamma-ray) by $p-\gamma$ interactions and proton synchrotron emission has been studied 
before (\citealt{Bottcher et al. (2013)}, \citealt{Basumallick & Gupta (2016)}). Hadronic models require super-Eddington luminosities to explain
the gamma-ray flux.

A two zone modeling was considered earlier by \citealt{Nalewajko et al. (2012)}  after including Herschel observations, \textit{Fermi}-LAT, Swift,
SMARTS and Submillimeter Array data for explaining the spectral and temporal features of activities of PKS 1510-089 in 2011. 
From March to August 2015 this source was again very active. Optical R-band monitoring with ATOM, supporting H.E.S.S. observations, detected 
very high flux of optical photons (\citealt{Zacharias et al. (2016)}). 

 Its enhanced activity in very high energy gamma-rays was also observed by MAGIC telescope (\citealt{Ahnen et al. (2017)}) in May, 2015.  
 In the middle of a long high state in optical and gamma-rays, for the first time they detected a fast variability in very high energy gamma-rays. 
 Their observation periods 
 MJD 57160-57161(Period A) and MJD 57164-57166 (Period B) overlap with one of the flares identified as flare-B (MJD 57150-57180) in our work. 
 They collected simultaneous data in radio, optical, UV, X-ray, and gamma-ray frequencies for multiwavelength modeling. They noted most of the flux 
 variation happened in \textit{Fermi}-LAT and MAGIC energy bands.
 
 The complex nature of 
 multiwavelength emission indicated a single zone model is not suitable for explaining the flares of PKS 1510-089. 
 In the present work several months of observational data have been studied for multiwavelength modeling of the high state of PKS 1510-089 in 2015.
 
The paper is organised as follows: in Section 2, we have provided the details about the multiwavelength data used in our study. In Section 3, we have presented
 our results, in Section 4, we have discussed our results and compared with the previous studies on this source.
 
 \section{Multiwavelength Data Analysis}
 The \textit{Fermi}-LAT and Swift-XRT/UVOT observations together cover optical, ultraviolet, X-ray and gamma-ray energy bands
 which allow us to do multiwavelength variability analysis and modeling of blazar flares.

\subsection{\textit{Fermi}-LAT}
After the successful launch of \textit{Fermi} Gamma-ray Space Telescope in 2008, thousands of sources have been 
revealed in the gamma-ray sky by the on-board instrument, Large Area Telescope (LAT), in the past eleven years.
With a field of view of about 2.4 sr  (\citealt{Atwood et al. (2009)}) LAT covers 20\% of the sky at any time and scans the whole sky every three hours.
It is sensitive to photons having energy between 20 MeV to higher than 500 GeV.
The third Fermi source catalog (3FGL; \citealt{Acero et al. (2015)}) shows that the extragalactic sky is dominated by active galactic nuclei (AGN) emitting high energy 
gamma-rays. The FSRQ PKS 1510-089 has been continuously monitored by \textit{Fermi}-LAT since August 2008. We collected the data for the year
2015 and analyzed it for energy range 0.1--300 GeV. A circular region of radius 10$\degree$ is chosen around the source of interest and the 
circular region is known as region of interest (ROI). 
The detailed procedure to analyze the \textit{Fermi}-LAT data is given in \citet{Prince et al. (2018)}. 

The data analysis also takes care of contamination from Earth's limb gamma-rays by rejecting the events having zenith angle higher than 90$\degree$.
In this analysis, we have used the latest instrument Response Function ``P8R2$\_$SOURCE$\_$V6" provided by the Fermi Science Tools.
\begin{figure*}
\begin{center}
 \includegraphics[scale=0.45]{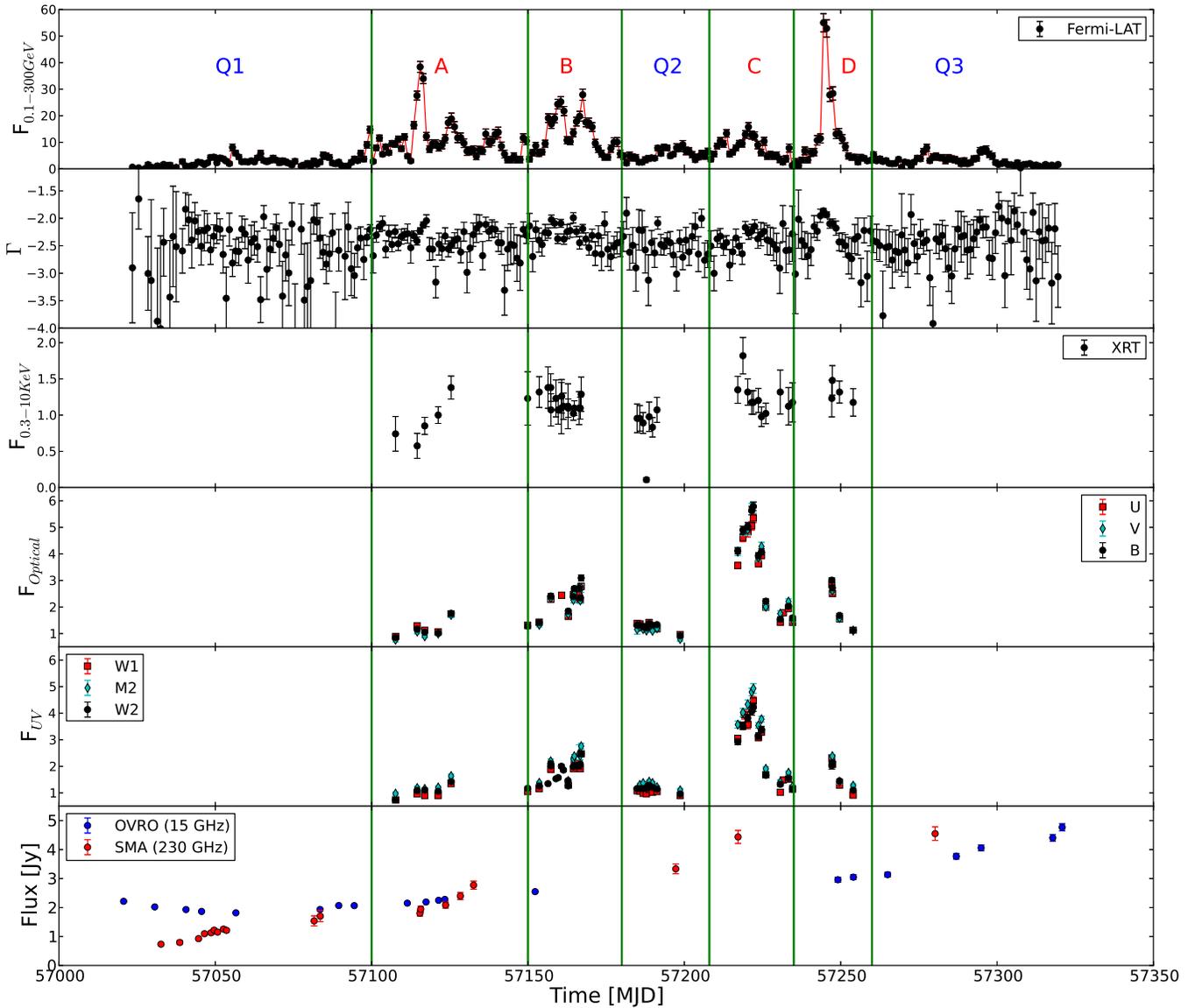}
\end{center} 
\vspace{-20pt}
 \caption{Light curve of PKS 1510-089 during 2015. Four flares A, B, C, and D have been detected with three quiescent states Q1, Q2, and Q3.
 Vertical green lines separate the different states of the source. Top panel represents the \textit{Fermi}-LAT data for 1 day binning along with 
 corresponding photon spectral index in second panel. Swift-XRT and UVOT light curves are shown in panel 3rd, 4th, and 5th.
 The last panel shows the radio light curve in two different
 energy bands, 15 and 230 GHz. The $\gamma$-rays flux data points are in units of 10$^{-7}$ ph cm$^{-2}$ s$^{-1}$ and X-ray/UV/Optical are
in units of 10$^{-11}$ erg cm$^{-2}$ s$^{-1}$.}
 \end{figure*}
%\begin{figure*}
%\centering
% \includegraphics[scale=0.45]{BVRJK.eps}
 %\includegraphics[scale=0.45]{radio.eps}
% \caption{Smarts B, V, R, J, and K band light curve for PKS 1510-089 during flare of 2015. After MJD 57204 (June 2015) Smarts has not 
% observed the source but it again started observing PKS 1510-089 from March 2016.}
%\end{figure*} 

\subsection{Swift-XRT/UVOT}
Swift data for PKS 1510-089 has been collected from \textit{HEASARC} webpage\footnote{https://heasarc.gsfc.nasa.gov/cgi-bin/W3Browse/w3browse.pl} for
a period of one year during 2015, which is part of archived data. 
In total 44 observations were reported during 2015 .
 A task \textit{`xrtpipeline'} version 0.13.2 has been run for every observation to get the cleaned event files.
The latest version of calibration files (CALDB version 20160609) and standard screening criteria have been used to re-process the raw data. 
Cleaned event files corresponding to the Photon Counting (PC) mode have been considered for further analysis. 
A circular region of radius 20 arc seconds around the source and away from the source has been chosen for the source and the background respectively
while analyzing the XRT data.
The X-ray light curve and spectra have been extracted by a tool called \textit{xselect}. 
The spectrum has been obtained and  fitted in ``xspec'' using simple power law model with the galactic absorption column density $n_H$ = 6.89$\times$10$^{20}$ 
cm$^{-2}$ \citep{Kalberla et al. (2005)}. 
The Swift Ultraviolet/Optical Telescope (UVOT, \citealt{Roming et al. (2005)}) also observed PKS 1510-089 in all the six filters U, V, B, W1, M2 and W2.
The source image has been extracted by choosing a circular region of 5 arc seconds around the source. 
Similarly, the background region has also been chosen with a radius of 10 arcseconds away from the source. 
The task `uvotsource' has been used to extract the source magnitudes and fluxes. 
Magnitudes are corrected for galactic extinction (E(B-V) = 0.087 mag; \citealt{Schlafly and Finkbeiner (2011)}) and converted into a flux
using the zero points (\citealt{Breeveld et al. (2011)}) and conversion factors (\citealt{Larionov et al. (2016)}). 

%\subsection{Smarts Observatory}
%PKS 1510-089 is also observed in different bands (Optical/IR) of Smarts Observatory\footnote{http://www.astro.yale.edu/smarts/glast/home.php} 
%as a part of \textit{Fermi} monitoring programme.
%The data reduction and analysis procedure are given in \citet{Bonning et al. (2012)}. We have collected the data for B, V, R, J, and K band for PKS 1510-089
%during a flare of 2015.

\subsection{Radio data at 15 and 230 GHz}
PKS 1510-089 was also observed in radio wavelength by OVRO \citep{Richards et al. (2011)}\footnote{http://www.astro.caltech.edu/ovroblazars/index.php?page=sourcelist} 
at 15 GHz and by Sub-millimeter array 
(SMA)\footnote{http://sma1.sma.hawaii.edu/callist/callist.html} at 230 GHz 
\citep{Gurwell et al. (2007)} as a part of the Fermi monitoring program. We have collected the data for the year 2015 from both the observatories.

\section{Results}
In this section, we have presented the results obtained from temporal and spectral analysis, and we have discussed the importance of these results
in multi-wavelength SED modeling.

\subsection{Multiwavelength Light Curves}
Multiwavelength light curves are shown in Figure 1, where they show indication of flares in various wavebands during the year of 2015 for 
PKS 1510-089.

The topmost panel of Figure 1 represents the gamma-ray light curve. 
The gamma-ray light curve is divided into different states on the basis of the fluxes observed during different time periods. We have also estimated the 
fractional rms variability amplitudes (\citealt{Fossati et al. (2000)}; \citealt{Vaughan et al. (2003)}) to identify the different states. If the 
value of the fractional variability during a time period is more than 0.5 (50$\%$) then it is considered as a flaring state.
During the time period MJD 57023--57100,
 the average flux of the source is found to be 2.49$\pm$0.10($\times$10$^{-7}$ ph cm$^{-2}$ s$^{-1}$) and it does not change significantly. 
 The fractional variability amplitude is found to be 0.50$\pm$0.04. This period has been identified as quiescent state Q1. 

The source started showing high activity
from MJD 57100 and continued for almost 50 days till MJD 57150. The average flux measured during this period is 10.06$\pm$0.19 ($\times$10$^{-7}$ ph cm$^{-2}$ s$^{-1}$)
which is five times higher than that in the quiescent state Q1. The fractional variability found during this period is 0.67$\pm$0.02 which confirms that the
source is more variable than in state Q1. This period is defined as flare A in our Figure 1. 

After the end of flare A, in 2-3 days the
flux again started rising and it lasted for a month. This period is noted as MJD 57150--57180. The average flux estimated during
this period is 12.17$\pm$0.26 ($\times$10$^{-7}$ ph cm$^{-2}$ s$^{-1}$). The fractional variability measured during this period is 0.57$\pm$0.02, 
which is higher than that in the state Q1. This period of high flux and high variability amplitude is referred as flare B in our work. 

After flare B, the source flux
became lower compared to flare A and B and  the source remained for a month ( MJD 57180--57208) in this low state. The average flux obtained during this
low state is 5.37$\pm$0.21 ($\times$10$^{-7}$ ph cm$^{-2}$ s$^{-1}$) and the fractional variability amplitude measured as 0.28$\pm$0.04. The
flux variability is not significant during this period as it is below 30$\%$. We have named this period as quiescent state Q2.

During the time period MJD 57208--57235, the source again went to a high state compared to Q1 and Q2, with average flux 
7.20$\pm$0.23($\times$10$^{-7}$ ph cm$^{-2}$ s$^{-1}$). The flux variability amplitude measured during this period is 0.51$\pm$0.03, which is just
above the limit we have set (50$\%$). The large average flux and 51$\%$ fractional variability suggest that this period is different from 
the state Q1 and Q2. Therefore, this state is identified as flare-C in gamma-ray which also have a strong flaring counter-part in optical/UV.

%This state identified as flare-C, which is more like an orphan optical/UV flare.

A higher state surpassing all the observed flares and quiescent states was observed during the time period MJD 57235--57260. The average
flux estimated during this period is 12.67$\pm$0.36 ($\times$10$^{-7}$ ph cm$^{-2}$ s$^{-1}$), which is much higher than the average flux observed
during any of the other states. A huge flux variation can be seen from Figure 1, and the fractional variability amplitude measured during this period
is 1.16$\pm$0.03 ($>$ 100$\%$). This period of high state is defined as flare D. 

After flare D, the flux decreased sharply and 
attained an average value of 2.93$\pm$0.12 ($\times$10$^{-7}$ ph cm$^{-2}$ s$^{-1}$). The source continued to have this average flux over a long period
of time from MJD 57260--57320. A small variation in flux was seen during this period (see Figure 1), for which the fractional variability amplitude has been
estimated as 0.50$\pm$0.05. Since the average flux in this state is very low compare to the flaring states, we have considered this period as quiescent state Q3.

The maximum flux observed during flare A, B, C, and D are 38.4, 27.92, 15.78, and 55.05 ($\times$10$^{-7}$ ph cm$^{-2}$ s$^{-1}$) at MJD 57115.5,
57167.5, 57220.5, and 57244.5 respectively. Flare D has been identified as the brightest gamma-ray flare of the year 2015. 
In Figure 1, the gamma-ray light curve is binned in one day time bin. The other light curves do not have an equally spaced binning because
 different observations were carried out at different times. We have estimated the average time between two consecutive observations for X-ray and optical/UV
light curves. In X-ray it is found to be 3.4 days and in optical/UV band it is estimated as 4.1, 4.3, 3.8, 3.8, 4.4, 3.5 days for filters B, V, U, W1, M2, W2
respectively.

\subsection{$\gamma$-ray variability}
%The multi-wavelength light curve of PKS 1510-089 during the flare of 2015 is shown in Figure 1. The uppermost panel represents the \textit{Fermi}-LAT 
%data in the energy range of 0.1-300 GeV. 
During 2015, the source was very active as it had been also seen earlier. The maximum flux attained at 
this time is (5.50$\pm$0.34)$\times$10$^{-6}$ ph cm$^{-2}$ s$^{-1}$ with photon spectral index 1.86 at MJD 57244.5.
The variability of the source can be seen from the gamma-ray light curve in Figure 1, which represents all the flares along with the photon 
spectral index in the second panel. 

It is seen that as the flux increases in gamma-ray the photon spectral index becomes
harder and harder.
The flux doubling/halving time is estimated during the flaring episodes by using the following equation (\citealt{Brown (2013)};
\citealt{Saito et al. (2013)}; \citealt{Paliya (2015)}), % \citet{Zhang et al. (1999)},}
\begin{equation}
F_2 = F_1.2^{({t_2-t_1})/\tau_d}
%t_{var} =  \frac{F_1 + F_2}{2} \frac{t_2 - t_1}{|F_2 - F_1|}
\end{equation}
where, F$_1$ and F$_2$ are the fluxes measured at two consecutive time t$_1$ and t$_2$, and $\tau_d$ represent the doubling
/halving times scale. One day binned gamma-ray light curve, shown in Figure 1, 
revealed the flux doubling time of 10.6 hr, when the source flux is changing from 1.15$\times$10$^{-6}$ to 5.50$\times$10$^{-6}$ between MJD 57243.5 
to MJD 57244.5. 

%\begin{figure*}
% \includegraphics[scale=0.5]{0.1-300GeV.eps}
% \caption{One day binned gamma-ray light curve with corresponding spectral index. }
%\end{figure*}

\subsection{X-ray variability}
The source is also followed by the Swift-XRT/UVOT telescope to unveil the behavior in X-ray, UV, and optical bands. In the third panel of Figure 1, we have
shown the X-ray light curve in the energy range of 0.3--10 keV. X-ray light curve is scanned by equation 1 and the flux doubling time is estimated for
consecutive time interval and it is found that the source is less variable in X-rays, moreover flaring states cannot be
clearly identified. The flux doubling time estimated by using equation 1 from X-ray light curve is 2.5 days for F$_1$ = 1.38$\times$10$^{-11}$ erg cm$^{-2}$ s$^{-1}$ at
t$_1$ = MJD 57156.41 and F$_2$ = 1.07$\times$10$^{-11}$ erg cm$^{-2}$ s$^{-1}$ at t$_2$ = MJD 57157.34.

\subsection{Optical and UV variability}
The Swift-UVOT light curve is plotted in the fourth and fifth panels of Figure 1. The source variability is significant during flare C while flare A, B, and D are less variable. 
In these three flares (A, B, $\&$ D) the variability of the source is constrained by the number of observations.
Equation 1 applied to the entire light curve of optical and UV band and the flux doubling times estimated  
for U, B, and V band light curves are 1.0, 0.7, and 1.1 days respectively (see Table 1).
Similar behaviour has also been seen in UV band (W1, M2, W2). The flux doubling
times estimated in these three bands of UV (W1, M2, W2) are 0.8, 1.4, and 1.1 days.\\ %, which are comparable to each other. \\

%\begin{figure*}
%\includegraphics[scale=0.5]{radio.eps}
 %\includegraphics[scale=0.5]{kband.eps}
% \caption{OVRO (15 GHz) and SMA (230 GHz) radio light curve for PKS 1510-089 during flares of 2015.}
%\end{figure*}

\subsection{Radio Light curves}
The last panel of Figure 1 represents the radio light curve in two different frequencies.
Owens Valley Radio Observatory (OVRO) and Sub-millimeter array (SMA) telescope radio data at 15 GHz and 230 GHz are plotted in the last panel of 
Figure 1. The radio light curves during 2015 clearly show that the radio fluxes in both the bands are increasing towards the end of the year.
The maximum radio flux in 2015 has been recorded as 4.77 Jy and 4.55 Jy at 15 GHz and 230 GHz respectively. The flux doubling time 
for the radio light curve is not estimated because of the poorly sparse data points.\\
\\
The fractional variability (F$_{\rm var}$) in different wavebands are also estimated following the \citep{Vaughan et al. (2003)}. For the gamma-ray 
light curve F$_{\rm var}$ is found to be 1.04$\pm$0.01, which corresponds to more than 100$\%$ variability. The F$_{\rm var}$ estimated for X-ray 
light curve is 0.14$\pm$0.04, which is the lowest among all the wavebands. A good amount of fractional variability is noticed from the optical light
curve shown in Figure 1. The F$_{\rm var}$ in optical U, B, and V bands are found to be 0.55$\pm$0.01, 0.59$\pm$0.01, and 0.64$\pm$0.01 respectively, 
and in UV bands for W1, M2, $\&$ W2 filters it is found to be 0.56$\pm$0.01, 0.53$\pm$0.01, $\&$ 0.48$\pm$0.01.
The fractional variability is also computed for the radio light curve at 15 GHz and 230 GHz. From OVRO light curve at 15 GHz the F$_{\rm var}$ is 
found to be 0.34$\pm$0.01. The SMA light curve shown in Figure 1 shows the large fractional variability compared to the OVRO light curve and the
F$_{\rm var}$ is noticed as 0.60$\pm$0.02. 

%A trend of increasing F$_{\rm var}$ with energy is observed, which suggest that a large number of
%articles produces high energy emission. A similar trend is also seen for other FSRQs like Ton 599 (\citealt{Prince (2019)}) and CTA 102 
%(\citealt{Kaur and Baliyan (2018)}). An opposite trend is also noticed by \citep{Bonning et al. (2009)}, where they have compared the fractional
%variability for IR, optical, and UV bands and they found that F$_{var}$ decreases towards higher energy, which suggests the presence of steady 
%thermal emission from an accretion disk.

\begin{table}
\centering
\caption{We have scanned all the light curves shown in Figure 1 by equation 1 and the flux doubling times ($\tau_d$)  are estimated
for all the different bands. The units of gamma-ray fluxes (F$_1$ $\&$ F$_2$)
are in 10$^{-7}$ ph cm$^{-2}$ s$^{-1}$ and X-rays/Optical/UV are is units of 10$^{-11}$ erg cm$^{-2}$ s$^{-1}$.}
 \begin{tabular}{cccccc p{1cm}}
 \hline \\
 Telescope/Bands & F$_1$ & F$_2$ & t$_1$  & t$_2$  & $\tau_d$ \\
 \hline \\
 Fermi-LAT &&&& & (hr) \\
 $\gamma$-rays & 1.15  & 5.50 & 57243.5 & 57244.5 & 10.6    \\
 \hline \\
 Swift-XRT/UVOT &&&&&  (days)\\
 X-rays  &  1.38  &  1.07  & 57156.41  & 57157.34 & 2.5   \\
 U	& 2.27   & 2.78  & 57166.72  & 57167.01    & 1.0    \\
 B	& 2.35   & 3.09  & 57166.72  & 57167.01    & 0.7    \\
 V	& 2.24   & 2.70  & 57166.72  & 57167.01    & 1.1    \\
 W1	& 1.91   & 2.47 & 57166.72  & 57167.01     & 0.8   \\
 M2	& 2.40   & 2.76 & 57166.72  & 57167.01     & 1.4   \\
 W2	& 2.07   & 2.47 & 57166.72  & 57167.01     & 1.1   \\
 \hline\\
 %Smarts &&&&& \\
 %B	& 2.40   & 3.23 & 57156.19  & 57157.08   & 2.98  \\
 %V	& 3.07   & 4.14 & 57156.19  & 57157.09	& 2.99   \\
 %R	& 1.90   & 2.68 & 57113.28  & 57114.24	& 2.81   \\
 %J	& 3.53   & 5.20 & 57113.28  & 57114.24  & 2.50   \\
 %K	& 16.66   & 24.60 & 57185.18  & 57186.19  & 2.63  \\
 %\hline
 \end{tabular}
 \end{table}
 
\subsection{Cross-Correlation}
A cross-correlation study between different energy bands can be done to find out the location of different emission regions responsible for
multi-wavelength emission along the jet axis. The Discrete Correlation Function (DCF) formulated by \citet{Edelson and Krolik (1988)} 
can be used to estimate cross and auto-correlations of the unevenly sampled light curves. 
We have made a few different combinations to show the DCFs. The combinations are $\gamma$-Swift B band, $\gamma$-X-rays, 
$\gamma$-OVRO, $\gamma$-SMA, and OVRO-SMA. DCFs for all these combinations are shown in Figure 2. When the two light curves LC1 and LC2 are cross-correlated,
 a positive time lag between them implies that the light curve LC1 is leading with respect to LC2, and a negative time lag implies the opposite.
\begin{figure*}
 \begin{center}
 \includegraphics[scale=0.28]{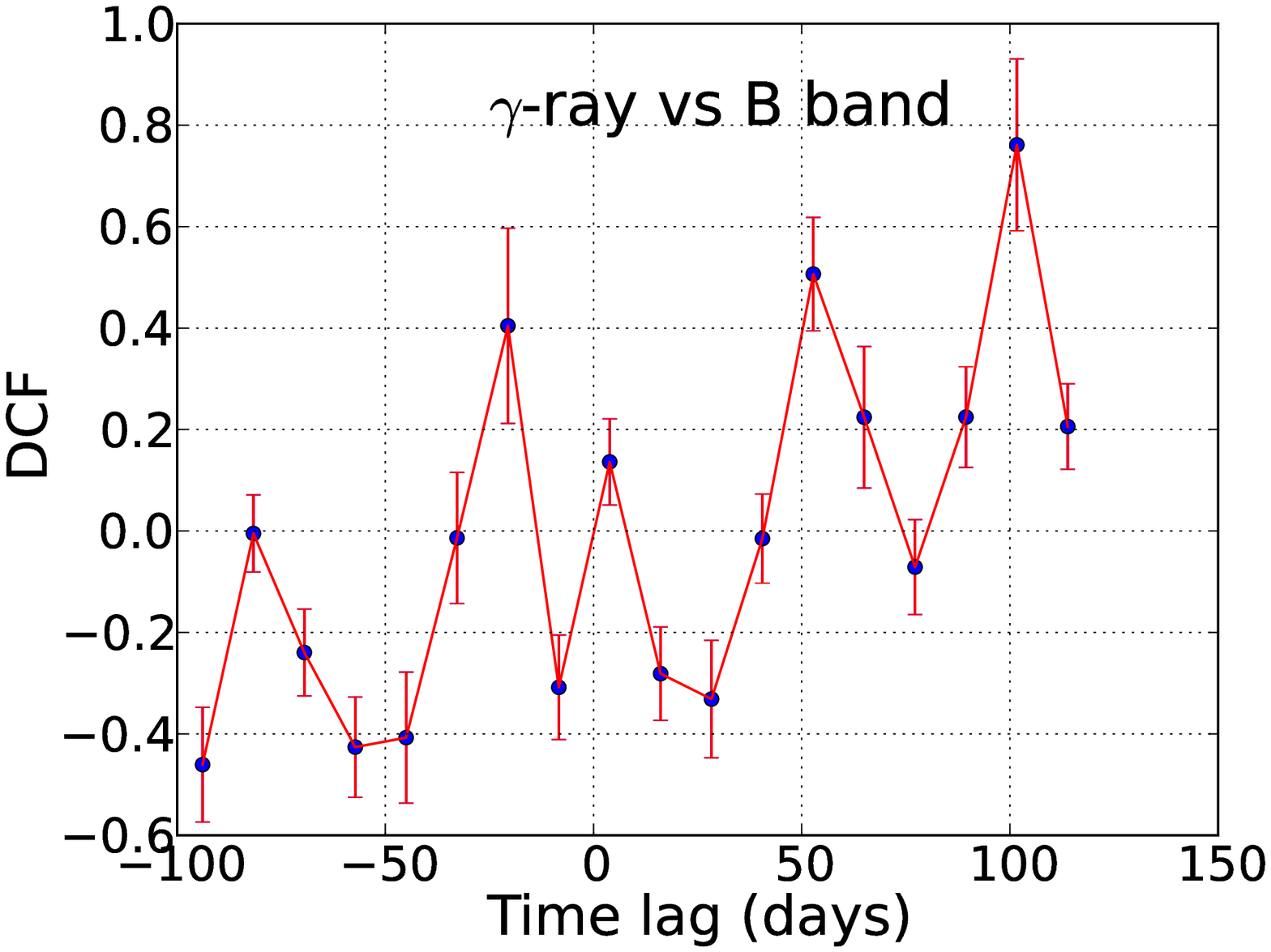}
  \includegraphics[scale=0.28]{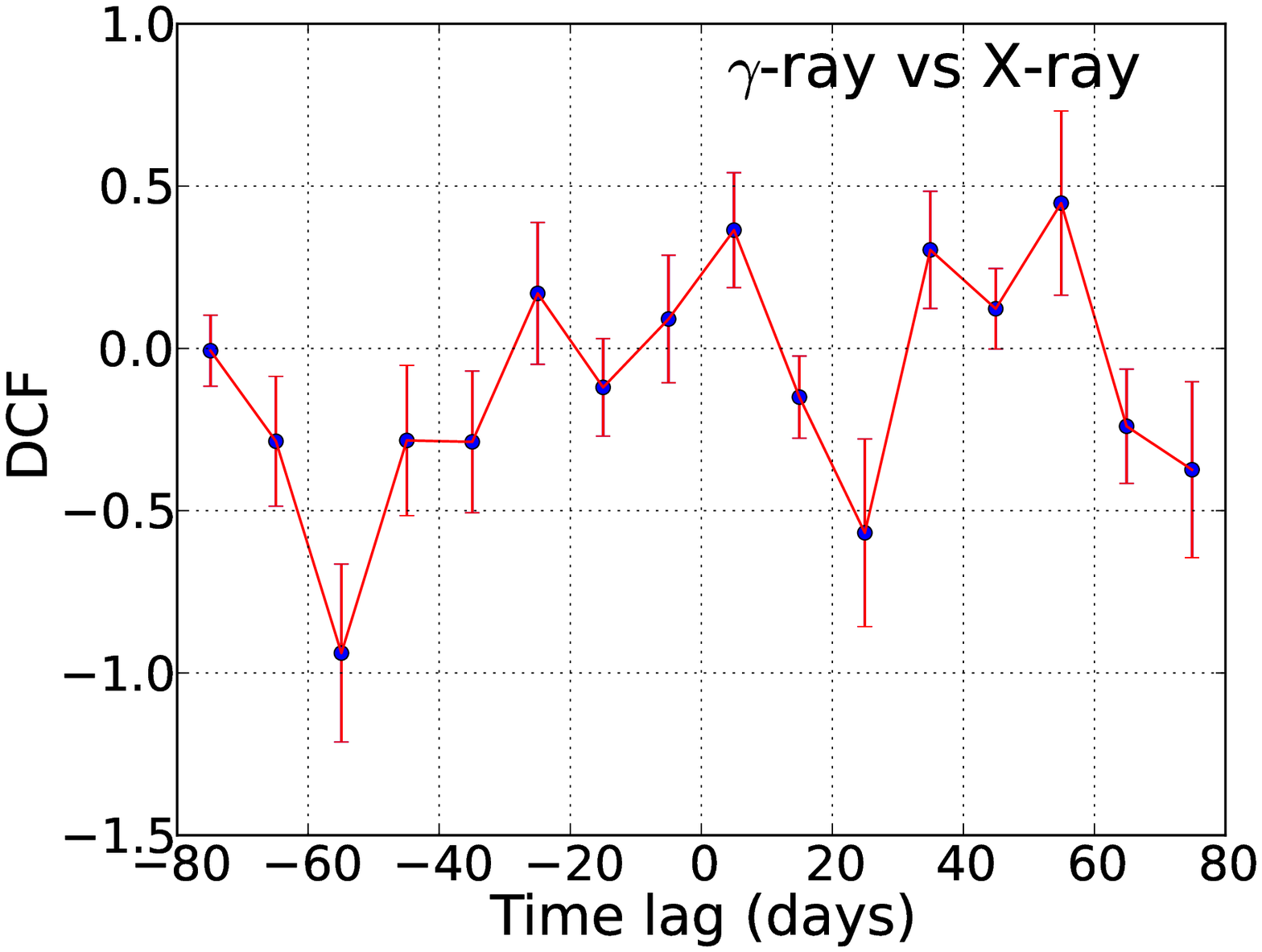}
  \includegraphics[scale=0.28]{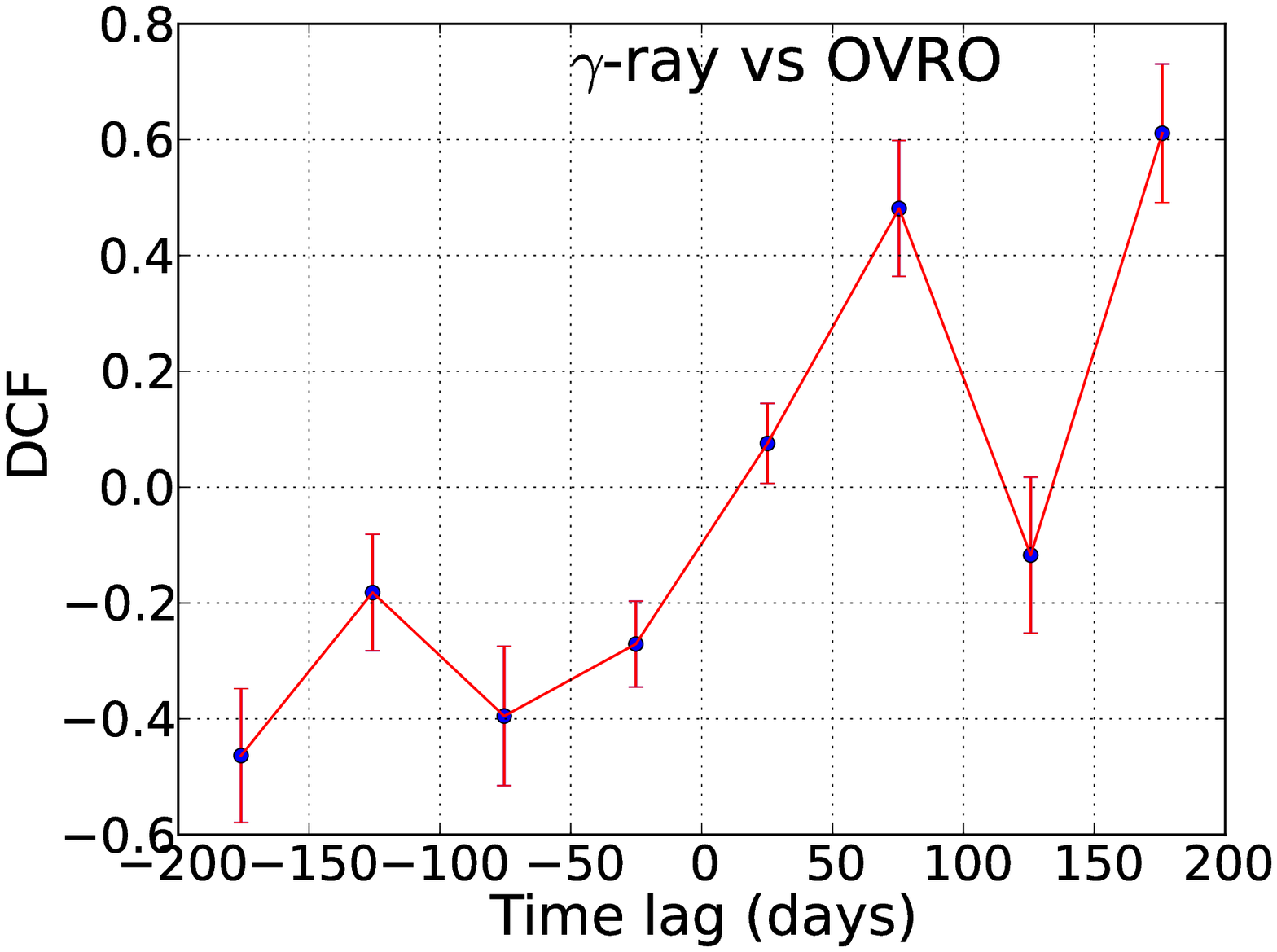}
  \includegraphics[scale=0.28]{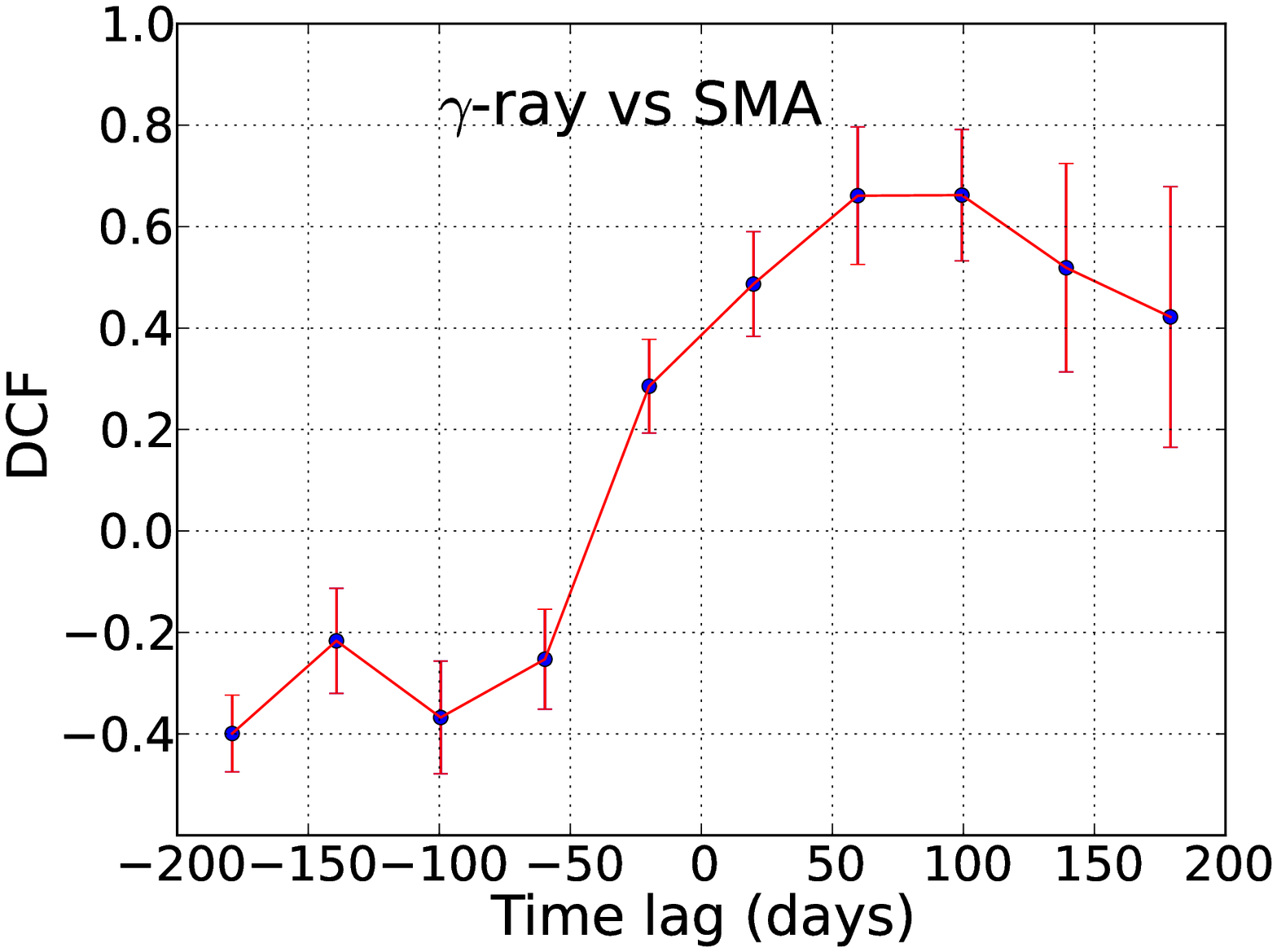}
  \includegraphics[scale=0.28]{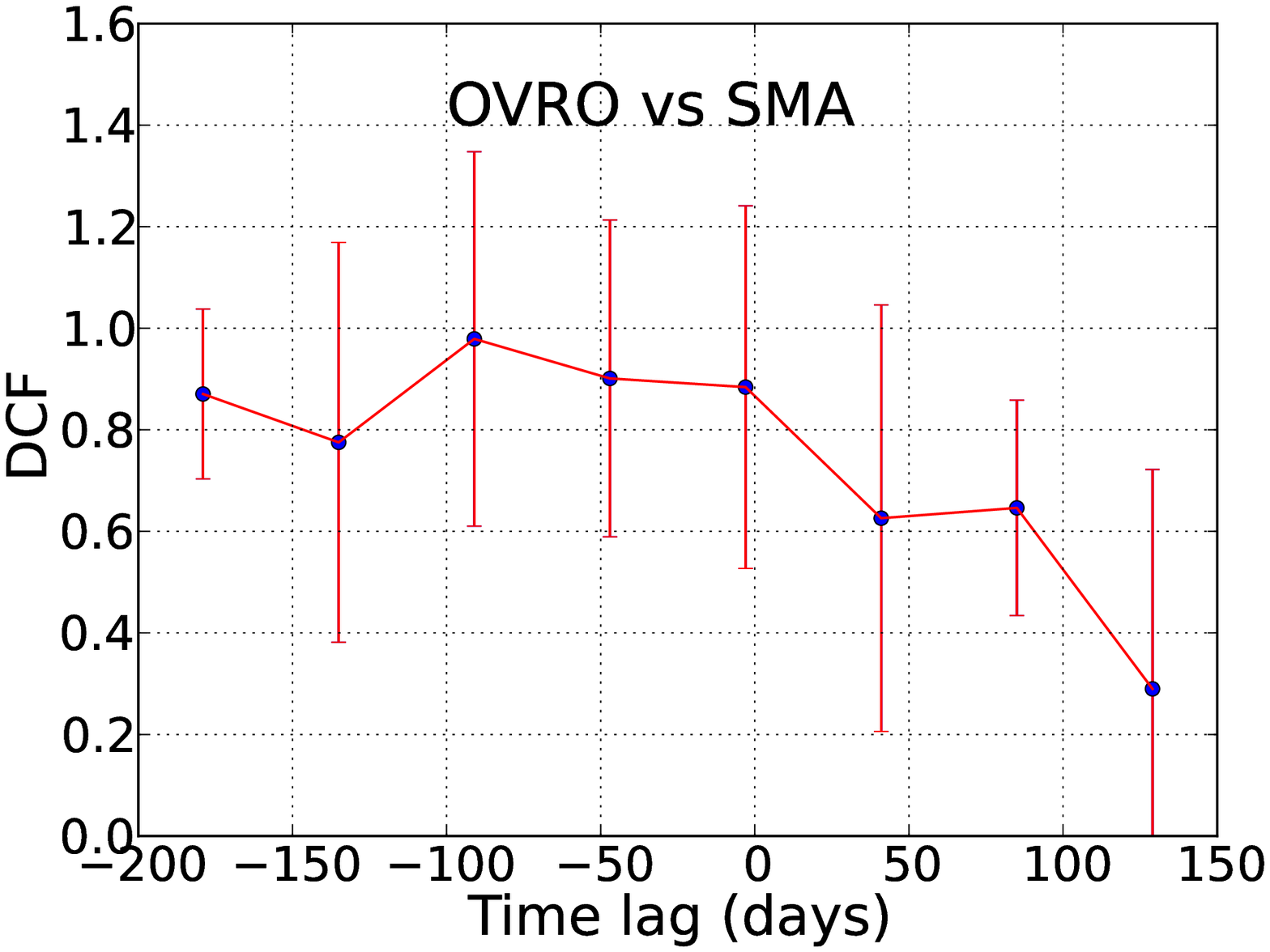}
  \end{center}
  \caption{DCFs for different combinations are plotted from top to bottom. The meaning of positive and negative time lags are described in section
  3.6.}
\end{figure*} \\
\\
\textbf{$\gamma$-ray vs optical B-band DCF:} \\
\\
The left most plot of Figure 2 upper panel shows the DCF between $\gamma$-ray and Swift optical B-band, and it is found that there are different peaks at different time
lags. We select the peak near zero time lag to constrain the location of the emission region. The peaks at +52 days and -20 days and the other two outer peaks
could be due to strong gamma-ray flare correlating with strong optical flare within the total period used for DCF analysis. 
A peak is observed at time lag 3.9 days, though the correlation coefficient is not much significant. We have estimated the average time resolution
of the worst light curve and the DCF time bin is chosen as three times of this average time resolution (\citealt{Edelson and Krolik (1988)}; \citealt{Castignani et al. (2014)}). 
In case of gamma-ray vs optical B-band the
DCF time bin is 12.2 days. The peak found within the DCF time bin is not considered as a time lag.
The multiple peaks in DCF are also observed by \citet{Kushwaha et al. (2017)} for 3C 454.3 during segment 4 as mentioned in their paper. 
The zero time lag observed by \citet{Castignani et al. (2017)} for PKS 1510-089 and the
small time lag observed in our case are consistent with the results obtained by \citet{Abdo et al. (2010)}; and \citet{Nalewajko et al. (2012)}
for other epochs. A zero or small time lag between two different
emissions suggests their co-spatial origin. Similar results were also found for different sources (\citealt{Prince (2019)}, \citealt{Kaur and Baliyan (2018)}).
The inference of a small time lag between gamma-ray and optical B band emission has been used to assume that the gamma-ray and optical photons are produced in the same region 
by the inverse Compton and synchrotron emission of the same population of electrons respectively. \\
\\ \\
\textbf{$\gamma$-ray vs X-ray DCF:} \\ 
\\ 
The gamma-ray vs. X-ray DCF is shown in the middle plot of the upper panel of Figure 2. A peak is observed at time lag 4.99 days with a correlation coefficient 
0.36$\pm$0.17. The DCF time bin 10.2 days is chosen on the basis of the average time resolution of the X-ray light curve. The observed peak is within the
DCF time bin and hence is not considered as the time lag between gamma-ray and X-ray emission. The peak observed at the edge of the DCF can be discarded.
A time lag of 50 days between gamma-rays and X-rays has been seen by \citet{Castignani et al. (2017)} in PKS 1510-089. 
This time lag between gamma-rays and X-rays suggests that the X-rays might have been produced far away from the region of gamma-ray emission in the jet. 
%However, our DCF study does not show any such behavior. 
A small correlation coefficient found in our case makes our results consistent with the
result obtained by \citet{Abdo et al. (2010)}, where they have also not found any robust evidence of cross-correlation between gamma-ray and X-ray
at zero time lag. \\ \\
\textbf{$\gamma$-ray vs OVRO and SMA DCF:} \\ \\
The rightmost plot of the upper panel and left plot of the lower panel of Figure 2 represent the gamma-ray vs. OVRO (15 GHz) and gamma-ray vs. SMA (230 GHz) DCFs respectively.
In gamma-ray vs OVRO, a peak is observed at time lag 75 days which is almost equal to one third of the length of the OVRO light curve. 
Hence, it cannot be considered as a DCF peak. Similar behavior is also seen in gamma-ray vs SMA DCF, where a peak is observed at the time lag between
60--100 days. This peak also lies at one third of the length of the SMA light curve and hence cannot be considered as DCF peak. \\ \\
\textbf{OVRO vs SMA DCF:} \\ \\
We have also tried to estimate the DCF between OVRO (15 GHz) and SMA (230 GHz) and the result is shown in the lower panel of Figure 2.
The DCF analysis does not show any significant peak, hence it is difficult to comment anything about the correlation between these two emissions. 
\\ \\
From the DCF analysis, it is clear that no good correlation is observed in any of the pairs. One of the reasons behind this is non-availability
of good quality data and a significant number of observations in X-ray, optical, and radio wavelengths for this particular time of period.
Hence, it would not be justified to conclude anything about the locations of different emission regions from this analysis.

\subsection{Multiwavelength SED Modeling with GAMERA}
 Our analysis shows the source went in long and bright flaring episodes in 2015.
The four bright flares are identified as Flare-A, Flare-B, Flare-C, and Flare-D. The quiescent states (Q1 and Q3) were observed
before and after the flares and the quiescent state Q2 in between Flare-B and Flare-C.
We have produced the gamma-ray SED in the energy range 0.1--300 GeV, for all the four gamma-ray flares along with one of the quiescent states Q2,
by using the unbinned likelihood analysis. The observed gamma-ray spectrum 
are fitted with four different functional forms Power Law (PL), Log Parabola (LP), Broken Power Law (BPL) and Power Law with Exponential cut-off 
(PLEC) as discussed in \citet{Prince et al. (2018)}.
We have found that the gamma-ray SED data points for all the flares and state Q2 are well
described by log-parabola (LP) distribution function. A LP photon spectrum can be produced by radiative losses of a LP electron spectrum 
(\citealt{Massaro et al. (2004)}). Due to this reason we have considered LP distribution for the injected electron spectrum in our multiwavelength SED modeling.
In X-rays and UV/Optical, the SED data points are also produced. 
All the spectral data points are plotted together in Figure 3 and modeled using the publicly available
code GAMERA\footnote{http://libgamera.github.io/GAMERA/docs/main$\_$page.html} \citep{Prince et al. (2018)}. GAMERA solves the time-dependent 
transport equation for input injected electron spectrum, and it calculates the propagated electron 
spectrum, and further, it uses this propagated electron spectrum as an input and estimates the synchrotron, synchrotron self-Compton (SSC), and 
inverse-Compton (IC) emission. \\
The following continuity equation we have used in our study,
\begin{equation}\label{8}
\frac{\partial N(E,t)}{\partial t}=Q(E,t)-\frac{\partial}{\partial E}\Big(b(E,t) N(E,t)\Big)
\end{equation}
$Q(E,t)$ is the injected electron (electron and positron) spectrum, $N(E,t)$ is the propagated electron spectrum after the radiative loss, and $b(E,t)$ covers the 
energy loss rate of electrons due to the synchrotron, synchrotron self-Compton (SSC) and external Compton (EC) emission. 
The code GAMERA estimates the inverse Compton emission using the full Klein-Nishina cross-section from \citet{Blumenthal and Gould (1970)}.

The flux doubling times (see Table 1) estimated in different wavelengths
suggest different emission zones. The doubling times found in gamma-ray and UV/Optical bands 
are closer to each other which suggests they might have been produced in the same region. 
Two emission regions are considered in this work, one is responsible for optical/UV and gamma-ray emission and another for the X-ray emission.

The locations of the emitting blobs along the jet axis are estimated by using the flux doubling time scales. We have used the following relation,
\begin{equation}
d = \frac{c t_{d} \delta}{(1+z)\theta_{jet}}
\end{equation}
where $t_{d}$ is the flux doubling time and $\theta_{jet}$ is the half opening angle of the jet 
(\citealt{Kaur and Baliyan (2018)}), d is the distance of the emitting region from the central supermassive black holes (SMBH), $c$ is the 
speed of light in vacuum, $z$=0.361 is the redshift of the source, and $\delta$=25 is the Doppler factor.
The jet opening angle was estimated from the radio observations by using the relation $\theta_{jet} = \theta_p sin \langle \Theta_0 \rangle$,
where $\theta_p$ = 4.8$\degree$ is the projected half opening angle, and $\langle \Theta_0 \rangle$ is the angle between jet axis 
and the line of sight. With the values of $\theta_p$ and $\langle \Theta_0 \rangle$ from \citet{Jorstad et al. (2005)} the jet opening 
angle  is found to be 0.12$\degree$. 
The observed flux doubling times for gamma-ray and X-rays are 10.6 hr and 2.5 days respectively, which are used to estimate the distances of the 
emission regions by using equation 3. The distance of gamma-ray emitting blob from the central SMBH is estimated at 1.76$\times$10$^{17}$ cm 
and the location of the X-rays emitting blob is estimated as 1.0$\times$10$^{18}$ cm along the jet axis.
The exact boundary of the broad line region (BLR) is not known, but we have some idea about the
radius of the BLR. To estimate the size of BLR and Dusty torus (DT), a simple scaling law is given by \citet{Ghisellini and Tavecchio (2009)}.
It only depends on the disk luminosity (L$_{disk}$). The relations are R$_{BLR}$ = 10$^{17}$L$_{d,45}^{1/2}$ and R$_{DT}$ = 2.5$\times$10$^{18}$L$_{d,45}^{1/2}$,
where L$_{d,45}$ is the disk luminosity in units of 10$^{45}$ erg/s. 
The disk luminosity has been estimated earlier by several authors 
(\citealt{Celotti et al. (1997)}; \citealt{Nalewajko et al. (2012)}) in the range of 3 -- 7$\times$10$^{45}$ erg s$^{-1}$. 
Using the typical value of disk luminosity
(L$_{disk}$ = 6.7$\times$10$^{45}$ erg s$^{-1}$), we have found that the radius of BLR (R$_{BLR}$) is 2.6$\times$10$^{17}$ cm and the size of the DT (R$_{DT}$)  region
is 6.47$\times$10$^{18}$ cm. From this calculation we conclude that the gamma-ray emitting blob is located within the edge of the BLR whereas
the X-ray emitting blob lies outside the BLR, in the DT region. We use these inferences in the SED modeling with the time dependent code GAMERA.

All the different flares and quiescent state Q2 are modeled with GAMERA as shown in Figure 3. The model
parameters are presented in Table 2. 
The energy density of the external radiation field in the comoving jet frame is given as,
\begin{equation} \label{5}
 U'{_{ext}} = \frac{\Gamma^{2} {\xi_{ext}} L_{disk}} {4 \pi c R_{ext}^{2}}
\end{equation}
where ``ext" represents the BLR or DT. The values of $\xi_{BLR}$ = 0.06, and $\xi_{DT}$ = 0.12 are comparable to \citealt{Barnacka et al. (2014)} 
and the jet Lorentz
factor $\Gamma$ = 20, taken from \citet{Aleksic et al. (2014)}. Using equation 4, the BLR energy density in the jet comoving frame is estimated as
$U'{_{BLR}}$ = 6.41 erg cm$^{-3}$ and DT energy density as $U'{_{DT}}$ = 2$\times$10$^{-3}$ erg cm$^{-3}$. The temperature of the BLR is used from
\citet{Peterson (2006)}, T$_{BLR}$ = 10$^{4}$K and the temperature of the DT region, T$_{DT}$ = 10$^{3}$K from \citet{Ahnen et al. (2017)}.

The Doppler factor ($\delta$) and Lorentz factor ($\Gamma$) for PKS 1510-089 have been chosen from an earlier study by \citet{Aleksic et al. (2014)}.
The sizes of the gamma-ray and X-ray emitting blobs are estimated by the relation R$<$ c$\tau_d$ $\delta$/(1+z), where $\tau_d$ denotes the 
doubling time in two different bands. 
The sizes of the emitting blobs are found to be 2.1$\times$10$^{16}$ cm and 1.2$\times$10$^{17}$ cm for gamma-ray and X-ray emission respectively.

The electron spectra for all the flares and the quiescent state evolve with time as the electrons lose
energy radiatively by synchrotron and IC emission. 
The duration of each flare and the quiescent state are significantly longer than the cooling time scale of electrons, as a result the electron
spectra become steady in a short time. The total time duration of flare A, B, C, and D are 50 days, 30 days, 28 days and 25 days respectively and
the quiescent state Q2 lasted for 28 days.
The synchrotron emission depends on the strength of the magnetic field and the luminosity of the relativistic electrons. The EC emission depends
on the energy density and temperature of the external photons and also the luminosity of the relativistic electrons.
The synchrotron self Compton (SSC) emission depends on the energy density of the synchrotron photons, which depends on the size of the blob, 
magnetic field and luminosity of the relativistic electrons. The SSC emission is found to be very low in our model compared to the external 
Compton emission. For the given magnetic field in the DT region the synchrotron emission is found to be below 10$^{-13}$ erg/cm$^{2}$/s,
hence not visible in Figure 3.

The optical depth correction due to the absorption of gamma-rays by the EBL (extragalactic background light) is not important for the \textit{Fermi}-LAT 
observed gamma-ray flux from PKS 1510-089.
We have included the optical depth correction on the observed data points by MAGIC from \citealt{Ahnen et al. (2017)}. The de-absorbed data points 
are used in the SED modeling.
To obtain the best model fit to the data points we have optimized the following parameters e.g. the magnetic field in the blob, 
luminosity and spectral index of injected relativistic electrons, and also their minimum and maximum energies ($\gamma_{min}$, $\gamma_{max}$). 

We have assumed the ratio of pairs (electrons and positrons) to cold protons in the emission regions is $10:1$.
The jet power in the relativistic electrons and positrons, or the magnetic field, or the cold protons is calculated with this expression 
$\rm P_{e,B,p}=\pi \, r_{blob}^2 \, \Gamma^2 \, c \,U_{e,B,p}$, 
where $U_{e,B,p}$ denotes the energy density in electrons and positrons, or magnetic field, or cold protons. The total jet power is always found 
to be lower than the Eddington's luminosity of the source $6.86\times 10^{46}$ erg s$^{-1}$, calculated 
with the black hole mass given in \citealt{Abdo et al. (2010)}.
The parameter values which can explain the SEDs of the flares A, B, C, D, and 
the quiescent state Q2 are listed in Table 2. 

\section{Discussion}
Below we discuss about our results and compare them with those of previous work.
\subsection{Multi-wavelength Studies with SED Modeling}
The flaring states identified as A, B, C, D and the quiescent states Q1, Q2 and Q3 are shown in the gamma-ray light curve in Figure 1 along with the light curves in other
 wavelengths. The flux doubling times in different wavelengths are found to be different, which motivated us to fit the SED with two-zone model.
 The values of the parameters used in two zone modeling are displayed  in Table 2. The magnetic field in the first zone required to fit the optical 
 and UV data points is in the range of 2.8 to 5.1 Gauss. This emission zone is located near the outer boundary of the BLR region. The magnetic field 
 in the second zone located in the DT region is not constrained by optical or UV data in our model. It is assumed to be very low to minimize the jet power. 
 However, in principle, it could be higher. The X-ray flux constrains the jet power in electrons and positrons in the second zone. In the first zone
 the magnetic field has more jet power than that in electrons and positrons. 
 In this zone the electrons and positrons carry more energy during the flaring states. MAGIC detected very high energy gamma-rays 
 \citep{Ahnen et al. (2017)}
 during Flare B. The maximum energy of the relativistic electrons and positrons in our model is the highest during the flare B. In the 
 second zone also this jet power is expected to be higher during the flaring states if the X-ray flux is higher than that in the quiescent states.

\citealt{Abdo et al. (2010)} noted complex correlation between fluxes in different
 wavelengths during the flaring activity of PKS 1510-089 between September 2008 and June 2009.
The high state of PKS1510-089 in 2009 was also studied by \citealt{Marscher et al. (2010)}. They found that the gamma-ray peaks were simultaneous
with maxima in optical flux.

The 2009 GeV flares of PKS 1510-089 have been studied by \citealt{Dotson et al. (2015)}. They have discussed about the location  of these flares. 
 For two flares they have suggested that the emission region is at the DT region and for the other two at the vicinity of VLBI radio core.
\citealt{Barnacka et al. (2014)} modeled the high energy flares detected in March 2009 from PKS 1510-089. They have used the photons in the BLR
and DT regions for EC emission to model the flares. SSC emission is insignificant in their model. In their model the emission zone is located at
a distance of $7\times 10^{17}$ cm  from the black hole.

The low states of this source between 2012 to 2017 have been studied in detail recently using MAGIC data 
  (MAGIC Collaboration; \citealt{Acciari et al. (2018)}). 
Their analysis shows the location of the gamma-ray emission region is close to the outer edge of the BLR region. 
They have chosen two scenarios with the emission regions located at $7\times10^{17}$ cm and  $3\times10^{18}$ cm away from the black hole respectively. 
For the high state in 2015 \citealt{Ahnen et al. (2017)} located the emission region at $6\times10^{17}$ cm away 
from  the black hole. These estimates are comparable to our results. In their work, the gamma-ray emission region has a size of
 $2.8\times10^{16}$ cm, which is also comparable to the size of the gamma-ray emission region found in our study $2.1\times10^{16}$ cm.
 Our Lorentz factor and Doppler factor values are similar to \citealt{Aleksic et al. (2014)}. 
 
The light curves of PKS 1510-089 and  3C 454.3 were studied by  \citealt{Tavecchio et al. (2010)}  for the period from August 4, 2008 to January
31, 2010 to constrain the location of  emission region through rapid variability in gamma-rays in the Fermi-LAT data.
 From hour scale variability in gamma-ray flux they constrained the size of the emission region to be less than $4.8\times 10^{15}$ cm and  
 $3.5\times 10^{15}$ cm for PKS 1510-089 and 3C 454.3 respectively for Doppler factor $\delta=10$. Extreme value of Doppler factor $\delta=50$ 
 constrains the size of the emission region to less than 0.01 pc. They suggested such small emission regions are likely to be located near the 
 black hole. They concluded that the far dissipation scenario, where the gamma-ray emission region is located 10-20 pc away
  from the black hole is disfavoured. 
  
A time dependent modeling of gamma-ray flares of PKS 1510-089 has been carried out by \citealt{Saito et al. (2015)} within the framework of the internal shock scenario. 
They have shown that the emission region is located  between 0.3 pc to 3 pc from the black hole depending on whether the jet is freely expanding or 
collimated. They have discussed about non-uniformity of Doppler factor across the jet  due to the radial expansion of the outflow. This may result
in time distortion in the observed gamma-ray light curve, in particular, asymmetric flux profiles with extended decay time.

  The most variable blazar 3C 454.3 has been well studied and modeled with multiwavelength observations (\citealt{Finke (2016)}). 
  Multiwavelength temporal variability in 3C 454.3 during its active state in 2014 has been studied by  \citealt{Kushwaha et al. (2017)}. 
  They found in some of the epochs IR/optical and 
  gamma-ray fluxes show nearly simultaneous variation. Correlation in Optical and gamma-ray frequencies was observed in June, 2016 outburst 
  of 3C 454.3 (\citealt{Weaver et al. (2019)}).
  Recently, \citealt{Rajput et al. (2019)}  have 
  analysed quasi-simultaneous data at optical, UV, X-ray and gamma-ray energies collected over a period of 9 years, August 2008 to February 2017.
  They identified four epochs when the source showed large optical flares.
  The optical and gamma-ray flares are correlated in two epochs. In two other epochs the flares in gamma-rays are weak or absent.
  
 A correlation in optical and gamma-ray photons from flares of PKS 1510-089 during Jan 2009 to Jan 2010 has been suggested by \citealt{Castignani et al. (2017)}, 
 which could be a common feature among these blazars. 
 This inference has also been used in our analysis to model the SEDs. We also note that in some FSRQs like 3C 279 the time lag between optical 
 and gamma-ray emission could be due to the variations in the ratio of energy densities in external photon field and magnetic field with distance
 across the length of the jet (\citealt{Janiak et al. (2012)}).

\subsection{Gamma-Radio Correlation}
An interesting feature of these flares is the gradual increase in the radio flux over a long period of time. In the bottom panel of our Figure 1, the 
light curves at 15 GHz from OVRO observations and at 230 GHz from SMA observations are shown. DCF estimated between these two light curves does not
show any clear peak or lag in Figure 2.
Even at the end of the high state when the gamma-ray flux reached
the quiescent state Q3 the radio flux continued to increase. The OVRO flux reached the maximum level in October, 2016 and  subsequently,
decreased slowly.

\citealt{Ahnen et al. (2017)} also reported gradual increase in radio flux in the second half of 2015. They have shown the light curve of the 
radio core at 43 GHz.  A bright and slow radio knot K15 was ejected on MJD 57230$\pm$52. They associated the increase in radio flux with the 
ejection of the radio knot K15. Due to the large uncertainty in the ejection time of K15 it could not be associated with any particular GeV 
flare. 

A similar feature was also observed with the gamma-ray high state in Feb-April 2012 when a radio knot K12 emerged from the core 
(\citealt{Aleksic et al. (2014)}).
In the second half of 2011 PKS 1510-089 had a major outburst in radio flux. The outburst first peaked at higher frequency. The peak at 37 GHz 
was reached around 21 Oct, 2011 and later at 15 GHz around 15 Dec, 2011. After attaining the peaks the light curves gradually decayed. 
Small outbursts continued to happen after this. VLBA 43 GHz images show a new component (knot K11) in December 2011. This was also observed at 
15 GHz in MOJAVE as reported by \citealt{Orienti et al. (2013)}.
The temporal evolution of gamma-ray and radio flux suggests they are produced by different populations of electrons, located at different
regions along the length of the jet.

\section{Conclusions} 

In this work, the high state of PKS 1510-089 in 2015 has been studied using the gamma-ray data from Fermi-LAT, Swift XRT/UVOT and radio data 
from OVRO and SMA observatory. 
Four flares are identified as A (MJD 57100 to MJD 57150), B (MJD 57150 to MJD 57180), C (MJD 57208 to MJD 57235) and D (MJD 57235 to MJD 57260) 
between quiescent states Q1 and Q3. Between flare B and C a quiescent state Q2 (MJD 57180 to MJD 57208) has also been identified. The epochs of 
MAGIC observations of flares in 2015 are within the duration of our flare B. We have also included MAGIC data for this flare from \citealt{Ahnen et al. (2017)}. 
We have inferred about the locations of the emission regions in different wavelengths from the flux doubling time scales. 
It is found that the source is less variable in X-rays and the flaring states cannot be clearly identified.
In our work the optical and gamma-ray emission is assumed to be co-spatial. This region of emission is located within the edge of the BLR region and
the X-ray emission could be from the DT region. 
The modeling has been done with the publicly available time dependent code GAMERA considering two emission zones. A log parabola distribution
of injected electrons is propagated using the code GAMERA and subsequently, the synchrotron, EC and SSC emission has been calculated to fit the 
observed data.
The parameter values used in our two zone model are displayed in Table 2. The data could be adequately fitted by adjusting the injected electron 
spectrum and the magnetic field. The jet power required in this scenario is below the Eddington's luminosity of PKS 1510-089. \\ 

\textbf{Acknowledgements :}  The authors thanks the referee for fruitful comments to improve the paper.
This work has made use of publicly available \textit{Fermi}-LAT data from FSSC and XRT data analysis software 
(XRTDAS) developed by ASI science data center, Italy. Archival data from the SMA
observatory and from the OVRO 40-m monitoring programme \citep{Richards et al. (2011)} have also been used in this research. N. Gupta acknowledges
the hospitality of the Nicolaus Copernicus Astronomical Center. This work was partially supported by the Polish National Science Centre grant
2015/18/E/ST9/00580.

\begin{table*}
\centering
\caption{ Parameters of two zone modeling, injection spectrum of electrons dQ(E)/dE = Q$_{p}$ (E/E$_{p}$)$^{-\alpha-\beta \log(E/E_{p})}$
, E$_p$=90 MeV. 
%The jet power in cold proton is also estimated in few bright flares and for flare B, C, and D it is found to be 3.60$\times$10$^{45}$
%erg/s, 3.30$\times$10$^{45}$ erg/s, and 1.71$\times$10$^{45}$ erg/s respectively. 
}
\scalebox{1.2}{%
\begin{tabular}{c c c c c}
\toprule
Model parameters & BLR  & DT   \\
\midrule
Energy density in BLR/DT (erg/cm$^3$) & 6.41 & 0.002  \\

Temperature in BLR/DT (K) & 1e4  & 1e3  \\ 

Doppler factor ($\delta$) & 25   & 25  \\

Lorentz factor ($\Gamma$) & 20   & 20  \\

Size of blob (r$_{blob}$ cm) & 2.1$\times$10$^{16}$ & 1.2$\times$10$^{17}$ \\
\midrule 
Flare-A && \\
\midrule 
%Injected electron luminosity (L$_{inj}$ erg/sec) & 0.70$\times$10$^{42}$  & 1.74$\times$10$^{42}$  \\
jet power in electrons (P$_{e}$ erg/sec) & 1.15$\times$10$^{45}$ & 8.79$\times$10$^{44}$ \\

jet power in magnetic field (P$_{B}$ erg/sec) & 5.18$\times$10$^{45}$ & 2.16$\times$10$^{42}$ \\

Injected electron spectrum ($\alpha$) & 2.0  & 3.1  \\

$\beta$ & 0.08 & 0.08 \\

$\gamma_{min}$ & 100   &  25  \\

$\gamma_{max}$ & 9$\times$10$^{3}$   &  2.2$\times$10$^{3}$  \\

Magnetic field (B Gauss) & 2.8 & 0.01 \\
\midrule 
Flare-B && \\
\midrule 
%Injected electron luminosity (L$_{inj}$ erg/sec) & 1.43$\times$10$^{42}$  & 5.79$\times$10$^{42}$  \\
jet power in electrons (P$_{e}$ erg/sec) & 1.08$\times$10$^{45}$ & 1.67$\times$10$^{45}$ \\

jet power in magnetic field (P$_{B}$ erg/sec) & 8.10$\times$10$^{45}$ & 2.16$\times$10$^{42}$ \\

Injected electron spectrum ($\alpha$) & 2.1  & 3.3  \\

$\beta$ & 0.16 & 0.08 \\

$\gamma_{min}$ & 250   &  30  \\

$\gamma_{max}$ & 2.7$\times$10$^{4}$   &  4$\times$10$^{3}$  \\

Magnetic field (B Gauss) & 3.5 & 0.01 \\
\midrule
Flare-C && \\
\midrule
jet power in electrons (P$_{e}$ erg/sec) & 9.38$\times$10$^{44}$ & 1.30$\times$10$^{45}$ \\

jet power in magnetic field (P$_{B}$ erg/sec) & 1.72$\times$10$^{46}$ & 2.16$\times$10$^{42}$ \\

%Injected electron luminosity (L$_{inj}$ erg/sec) & 0.96$\times$10$^{42}$  & 6.35$\times$10$^{42}$ \\
Injected electron spectrum ($\alpha$) & 2.1  & 3.5  \\

$\beta$ & 0.17 & 0.10 \\

$\gamma_{min}$ & 190   &  24  \\

$\gamma_{max}$ & 8$\times$10$^{3}$   &  9$\times$10$^{2}$  \\

Magnetic field (B Gauss) & 5.1 & 0.01 \\
\midrule 
Flare-D && \\
\midrule
jet power in electrons (P$_{e}$ erg/sec) & 8.27$\times$10$^{44}$ & 2.51$\times$10$^{44}$ \\

jet power in magnetic field (P$_{B}$ erg/sec) & 7.65$\times$10$^{45}$ & 2.16$\times$10$^{42}$ \\

%Injected electron luminosity (L$_{inj}$ erg/sec) & 0.85$\times$10$^{42}$  & 5.79$\times$10$^{42}$  \\

Injected electron spectrum ($\alpha$) & 1.7  & 3.0  \\

$\beta$ & 0.07 & 0.07 \\

$\gamma_{min}$ & 170   &  25  \\

$\gamma_{max}$ & 1.2$\times$10$^{4}$   &  1.3$\times$10$^{3}$  \\

Magnetic field (B Gauss) & 3.4 & 0.01 \\
\midrule 
Quiescent State (Q2) && \\
\midrule
jet power in electrons (P$_{e}$ erg/sec) & 7.80$\times$10$^{44}$ & 1.63$\times$10$^{45}$ \\

jet power in magnetic field (P$_{B}$ erg/sec) & 9.55$\times$10$^{45}$ & 2.16$\times$10$^{42}$ \\
%Injected electron luminosity (L$_{inj}$ erg/sec)  & 0.70$\times$10$^{42}$  & 4.69$\times$10$^{42}$  \\
Injected electron spectrum ($\alpha$) & 2.2  & 3.3  \\

$\beta$ & 0.08 & 0.08 \\

$\gamma_{min}$ & 200   &  27  \\

$\gamma_{max}$ & 6$\times$10$^{3}$   &  8$\times$10$^{2}$ \\

Magnetic field (B Gauss) & 3.8 & 0.01 \\
\bottomrule
\end{tabular}
}
\label{Table:TB}
\end{table*}

\begin{figure*}
\centering
 \includegraphics[scale=0.40]{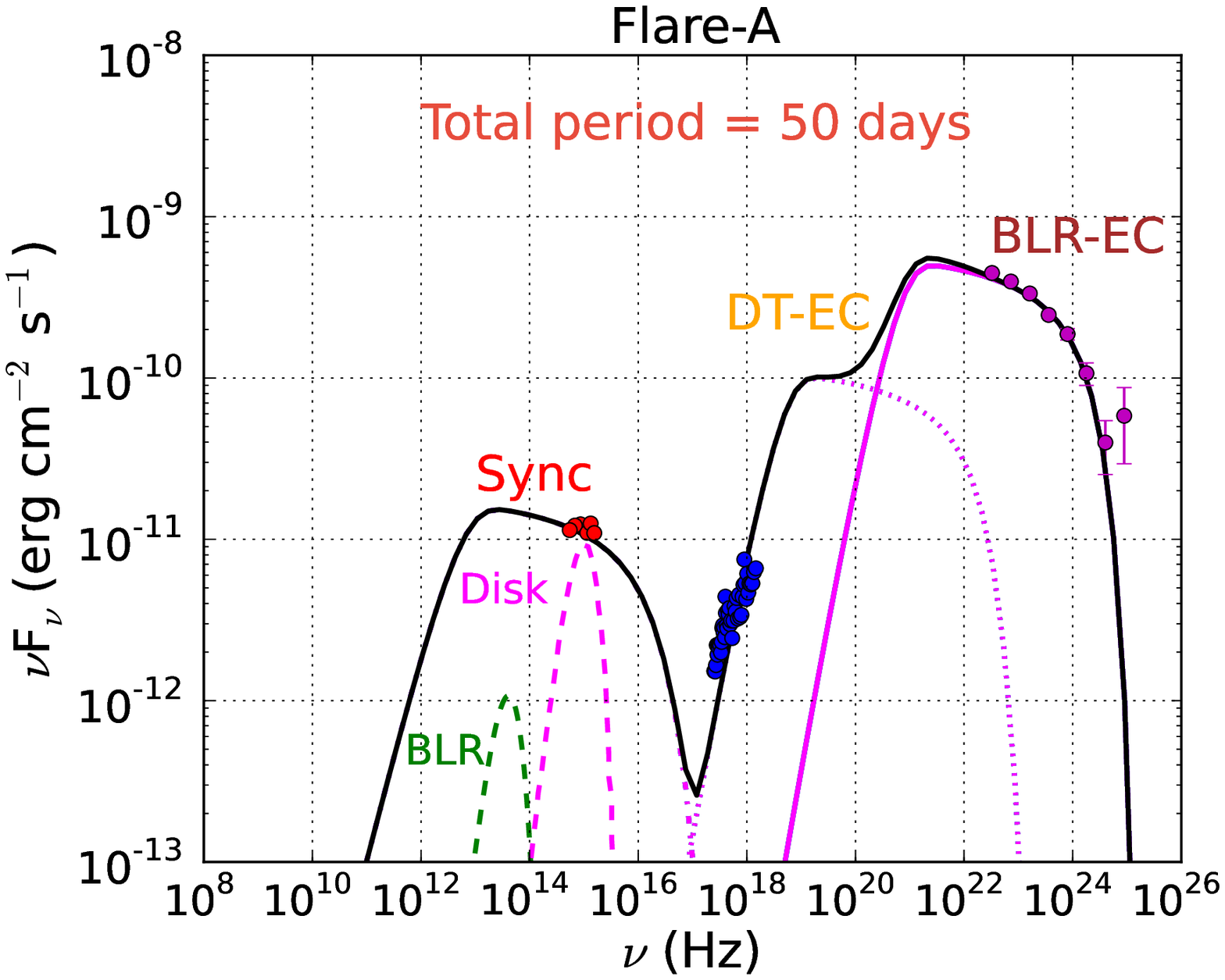}
 \includegraphics[scale=0.40]{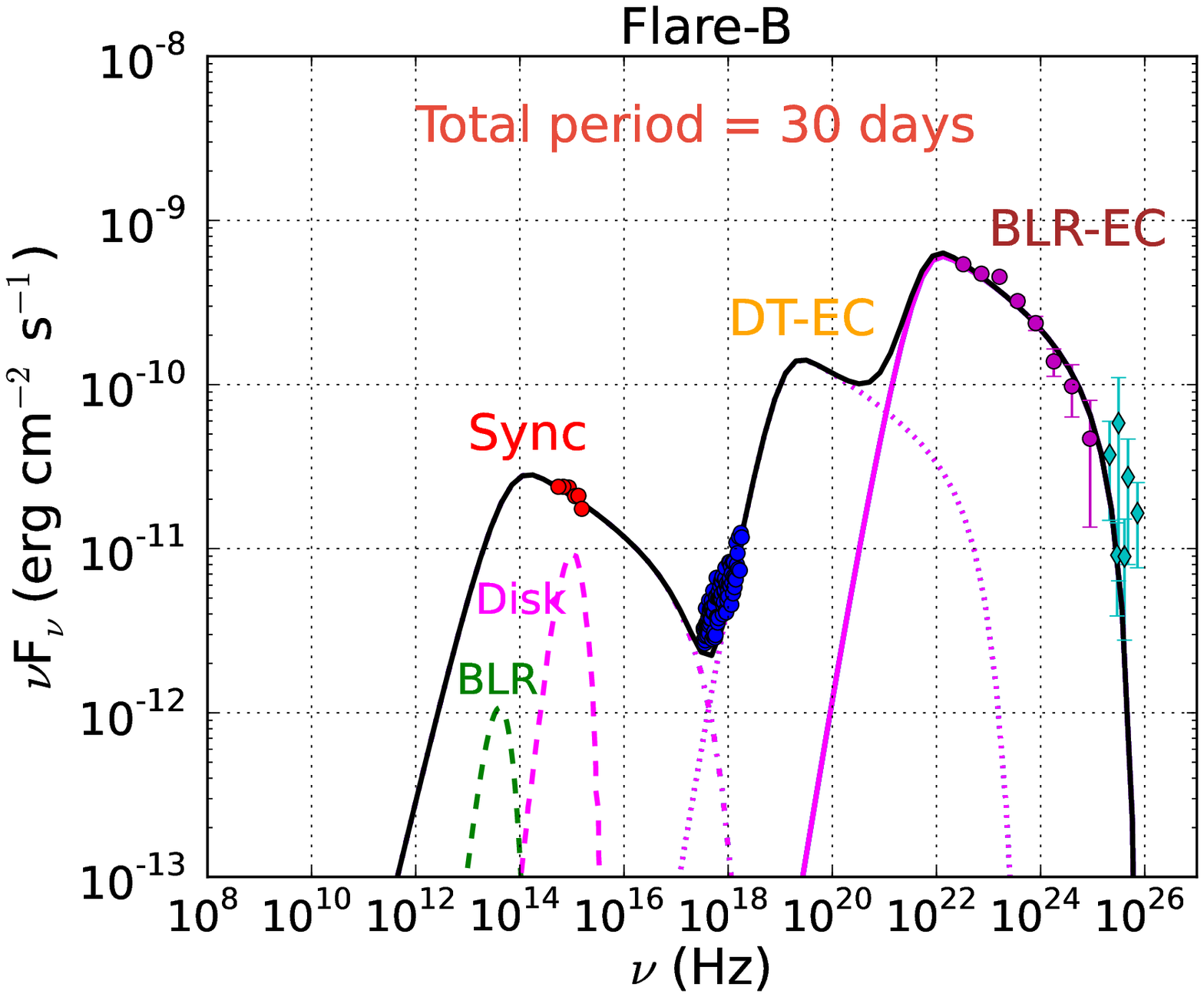} 
 \includegraphics[scale=0.40]{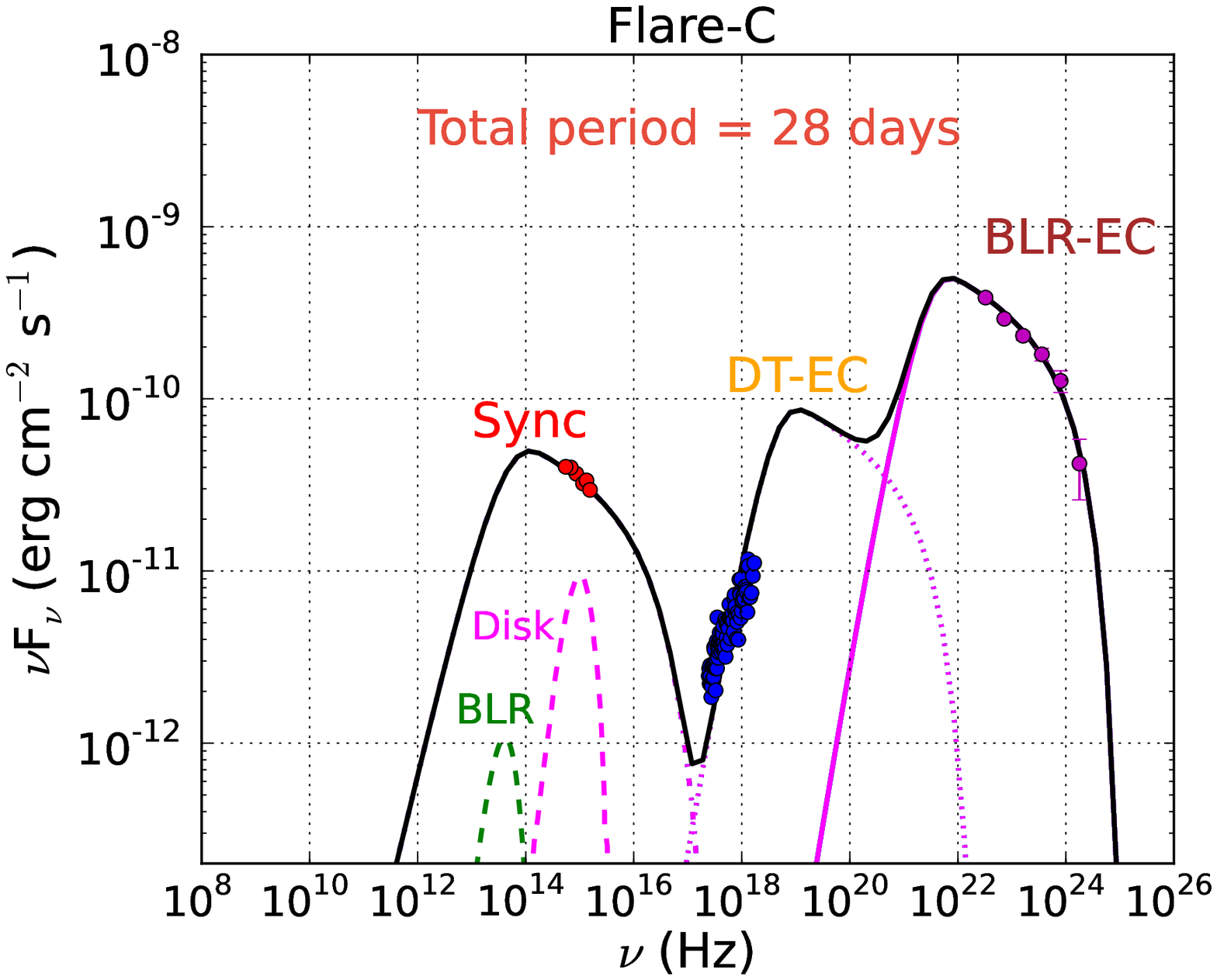}
 \includegraphics[scale=0.40]{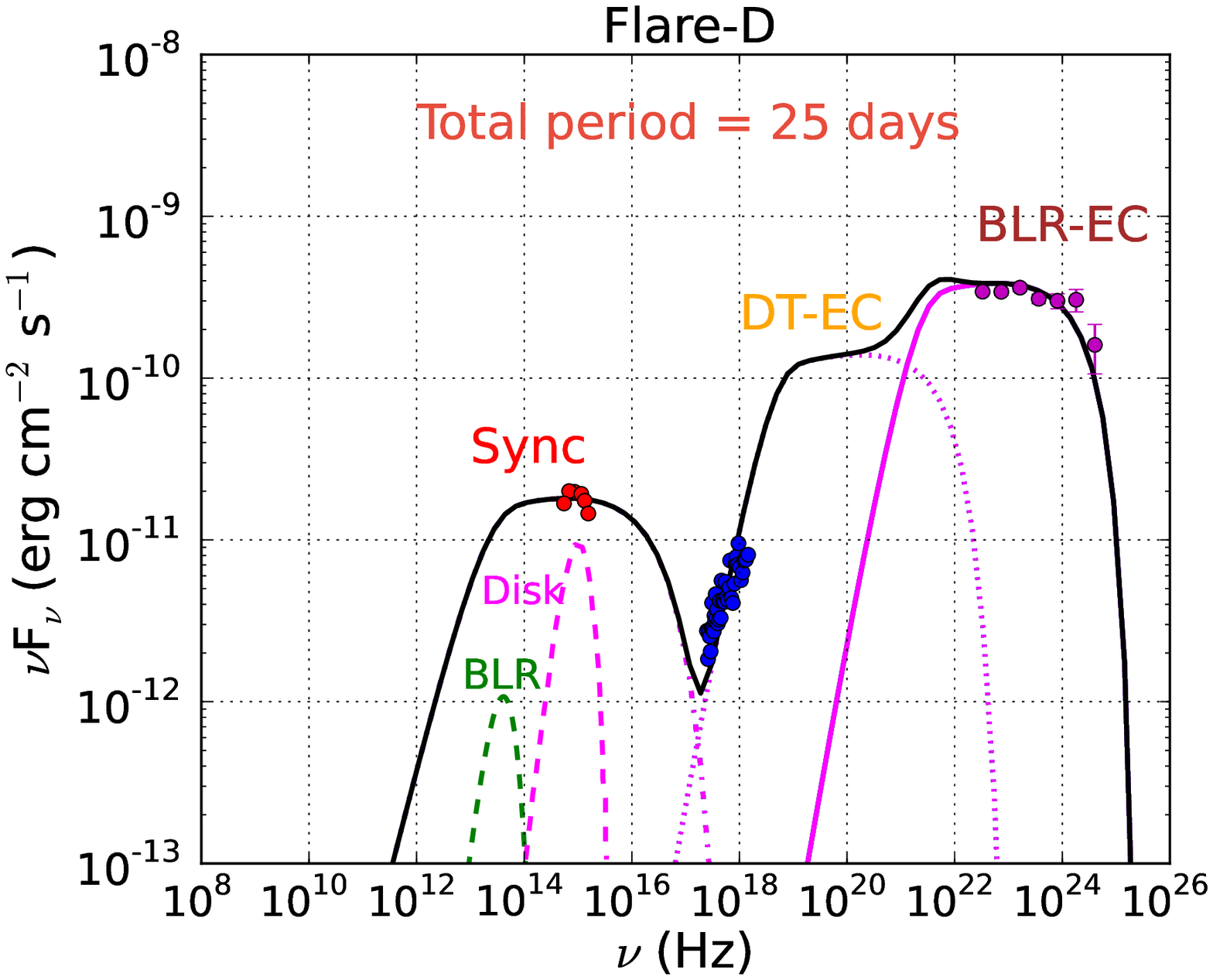}
 \includegraphics[scale=0.40]{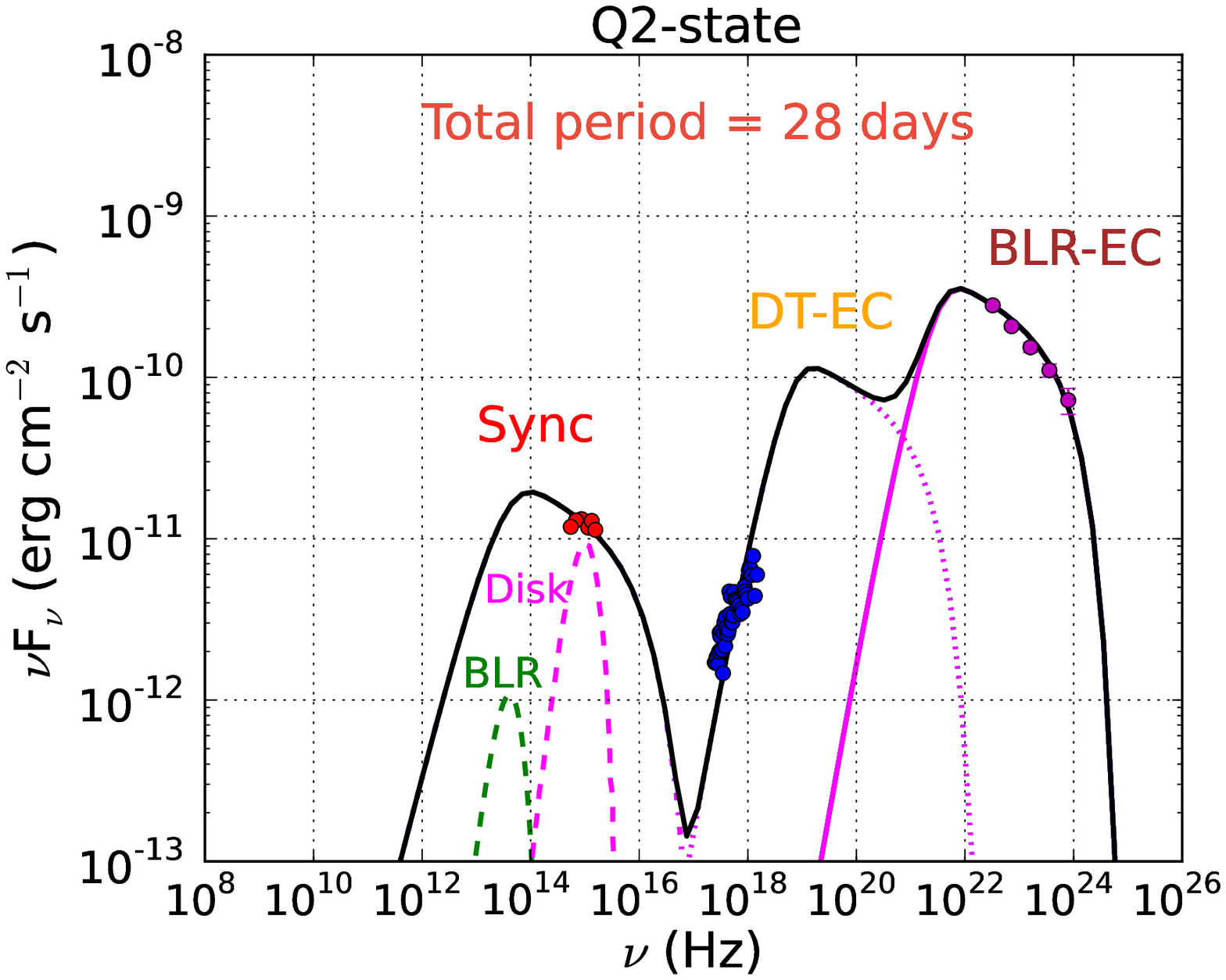}
 \caption{Multiwavelength SED modeling for four flares and quiescent state Q2. Swift XRT/UVOT data points shown in blue/red solid circles. 
 \textit{Fermi}-LAT data points shown in pink solid circles, for Flare B the de-absorbed MAGIC data points \citep{Ahnen et al. (2017)} shown with 
 teal diamonds. }
\end{figure*}

\bibliographystyle{plain}

\begin{thebibliography}{} 
\bibitem[Abdo et al. (2010)]{Abdo et al. (2010)} Abdo, A. A., Ackermann, M., et al.  2010, ApJ, 721, 1425
%\bibitem[ ]{×} Abdo, A. A., et al. 2010c, ApJ, 721, 1425
\bibitem[Acciari et al. (2018)]{Acciari et al. (2018)} Acciari, V. A. et al., 2018, A \& A, 619, 159
 %\bibitem[Abdo et al. (2010b)]{Abdo et al. (2010b)} Abdo, A. A., Ackermann, M., et al. 2010, ApJ, 722, 520
 \bibitem[Abramowski et al. (2013)]{Abramowski et al. (2013)}  Abramowski, A., et al.,  2013, A \& A, 550, 4 
 \bibitem[Aleksi{\'c} et al. (2014)]{Aleksic et al. (2014)} Aleksi{\'c}, J., Ansoldi, S., Antonelli, L.~A. et al. 2014, A\&A, 569, A46
 \bibitem[Atwood et al. (2009)]{Atwood et al. (2009)} Atwood, W.~B., Abdo, A.~A., et al., 2009, ApJ, 697, 1071
 \bibitem[Acero et al. (2015)]{Acero et al. (2015)} Acero, F., Ackermann, M., et al., 2015,  ApJS, 218, 23 
 \bibitem[Ahnen et al. (2017)]{Ahnen et al. (2017)} Ahnen, M.~L., Ansoldi, S., Antonelli, L.~A. et al. 2017, A \& A, 603, A29 
 \bibitem[Barnacka et al. (2014)]{Barnacka et al. (2014)} Barnacka, A., Moderski, R., Behera, B., Brun, P., Wagner, S., 2014, 567, A113
 \bibitem[Basumallick \& Gupta (2016)]{Basumallick & Gupta (2016)} Basumallick, P. P. \& Gupta, N., 2016, APh, 88, 1
 \bibitem[Blumenthal \& Gould (1970)]{Blumenthal and Gould (1970)}Blumenthal, R. \& Gould, G., 1970, Rev. Mod. Phys., 42, 237
 \bibitem[Bonning et al. (2009)]{Bonning et al. (2009)} Bonning, E, W., Bailyn, C., Urry, C. M., et al. 2009, ApJL, 697, L81
 \bibitem[B\"ottcher et al. (2013)]{Bottcher et al. (2013)} B\"ottcher M., Reimer A., Sweeney K. \& Prakash A. 2013, ApJ, 768, 54
 \bibitem[Breeveld et al. (2011)]{Breeveld et al. (2011)} Breeveld, A.~A., Landsman, W., et al., 2011, AIPC, 1358, 373-376
 \bibitem[Brown (2013)]{Brown (2013)} Brown, A., 2013, MNRAS, 431, 824-835
 \bibitem[Castignani et al. (2014)]{Castignani et al. (2014)} Castignani, G., Guetta, D., Pian, E., et al. 2014, A\&A, 565, A60
 \bibitem[Castignani et al. (2017)]{Castignani et al. (2017)} Castignani, G., Pian, E., Belloni, T M., et al. 2017, A\&A, 601, A30
 \bibitem[Celotti et al. (1997)]{Celotti et al. (1997)} Celotti, A., Padovani, P., \& Ghisellini, G. 1997, MNRAS, 286, 415
 \bibitem[Dotson et al. (2015)]{Dotson et al. (2015)} Dotson, A., Georganopoulos, M., et al., 2015, ApJ, 809, 164  
 \bibitem[Edelson \& Krolik (1988)]{Edelson and Krolik (1988)} Edelson, R. A., \& Krolik, J. H., 1988, ApJ, 333, 646
 \bibitem[Finke (2016)]{Finke (2016)} Finke, J. D., 2016, ApJ, 830, 94 
 \bibitem[Fossati et al. (2000)]{Fossati et al. (2000)} Fossati,G., Celotti, A., Chiaberge, M., et al. 2000, ApJ, 541, 153-165
 \bibitem[Fuhrmann et al. (2014)]{Fuhrmann et al. (2014)} Fuhrmann, L., Larsson, S., et al. 2014, MNRAS, 441, 1899
 \bibitem[Ghisellini \& Tavecchio (2009)]{Ghisellini and Tavecchio (2009)} Ghisellini, G., \& Tavecchio, F., 2009, MNRAS, 397,985
 \bibitem[Gurwell et al. (2007)]{Gurwell et al. (2007)} Gurwell, M. A., Peck, A. B., Hostler, S. R., Darrah, M. R., \& Katz, C. A. 2007, in 
	  Astronomical Society of the Pacific Conference Series, Vol. 375,From Z-Machines to ALMA: (Sub)Millimeter Spectroscopy of Galaxies, 
	  ed. A. J. Baker, J. Glenn, A. I. Harris, J. G. Mangum, \& M. S. Yun, 234
 \bibitem[Janiak et al. (2012)]{Janiak et al. (2012)} Janiak, M., Sikora, M., et al., 2012, ApJ, 760, 129  	  
 \bibitem[Jorstad et al. (2005)]{Jorstad et al. (2005)} Jorstad, S. G., Marscher, A. P., Lister, M. L., et al. 2005, AJ, 130, 1418
 \bibitem[Kalberla et al. (2005)]{Kalberla et al. (2005)} Kalberla, P.~M.~W., Burton, W.~B., et al., 2005, A \& A, 440, 775-782
 \bibitem[Kataoka et al. (2008)]{Kataoka et al. (2008)} Kataoka, J., et al., 2008, ApJ, 672, 787-799
 \bibitem[Kaur \& Baliyan (2018)]{Kaur and Baliyan (2018)} Kaur, N., Baliyan, K.~S., 2018, A \& A, 617, 59
 \bibitem[Kushwaha et al.(2017)]{Kushwaha et al. (2017)} Kushwaha, P., Gupta, A. C., Misra, R., Singh, K. P., 2017, MNRAS, 464, 2046
 \bibitem[Larionov et al. (2016)]{Larionov et al. (2016)} Larionov, V.~M., Villata, M., et al., 2016, MNRAS, 461, 3047-3056 
 \bibitem[Liodakis et al. (2017)]{Liodakis et al. (2017)} Liodakis, I., Marchili, N., et al. 2017, MNRAS, 466, 4625-4632
 \bibitem[Marscher et al. (2010)]{Marscher et al. (2010)} Marscher, A. P., Jorstad, S., Larionov, V., et al. 2010, ApJ, 710, L126 
 \bibitem[Massaro et al. (2004)]{Massaro et al. (2004)} Massaro, E.  Perri, M., Giomi, P. \& Nesci, R., 2004, A \& A, 413, 489
 \bibitem[Nalewajko et al. (2012)]{Nalewajko et al. (2012)} Nalewajko, K., Sikora, M., Madejeski, G., et al. 2012, ApJ, 760, 69
 \bibitem[Nalewajko (2013)]{Nalewajko (2013)} Nalewajko, K., 2013, MNRAS, 430, 1324-1333
 \bibitem[Orienti et al. (2013)]{Orienti et al. (2013)} Orienti, M. D., Ammando, F., Giroletti, M., et al. 2013, MNRAS, 428, 241
 \bibitem[Paliya (2015)]{Paliya (2015)} Paliya, V. S., 2015, ApJL, 808, L48
 \bibitem[Peterson (2006)]{Peterson (2006)} Peterson B. M., 2006 Alloin D. Lecture Notes in Physics, Vol. 693, Physics of Active Galactic Nuclei at 
    all Scales Springer Berlin 77
 \bibitem[Prince et al. (2017)]{Prince et al. (2017)} Prince, R., Majumdar P.,  Gupta N., 2017, ApJ, 844, 62 
 \bibitem[Prince et al. (2018)]{Prince et al. (2018)} Prince, R., Raman, G., et al. 2018, ApJ, 866, 16
 \bibitem[Prince (2019)]{Prince (2019)} Prince, R., 2019, ApJ, 871, 101
 \bibitem[Rajput et al. (2019)]{Rajput et al. (2019)} Rajput, B., Stalin, C. S. et al., 2019, MNRAS, 486, 1781 
 \bibitem[Richards et al. (2011)]{Richards et al. (2011)} Richards, J, L., Max-Moerbeck, W., et al. 2011, ApJS, 194, 29
 \bibitem[Roming et al. (2005)]{Roming et al. (2005)} Roming, P.~W.~A., Kennedy, T.~E., et al., 2005, ssr, 120, 95-142
 \bibitem[Schlafly \& Finkbeiner (2011)]{Schlafly and Finkbeiner (2011)} Schlafly, E.~F., Finkbeiner, D.~P., 2011, ApJ, 737, 103
 \bibitem[Saito et al. (2013)]{Saito et al. (2013)} Saito, S., Stawarz, L., Tanaka, Y. T., et al., 2013, ApJL, 766, L11
 \bibitem[Saito et al. (2015)]{Saito et al. (2015)} Saito, S., Stawarz, L., et al., 2015, ApJ, 809, 171
 \bibitem[Tanner et al. (1996)]{Tanner et al. (1996)} Tanner, A. M., Bechtold, J., Walker, C. E., Black, J. H., \& Cutri, R. M., 1996, AJ, 112, 62
 \bibitem[Tavecchio et al. (2010)]{Tavecchio et al. (2010)} Tavecchio, F., Ghisellini, G., et al.,  2010, MNRAS, 405, L94-L98
 \bibitem[Vaughan et al. (2003)]{Vaughan et al. (2003)} Vaughan, S., Edelson, R., Warwick, R. S., \& Uttley, P., 2003, MNRAS, 345, 1271
 \bibitem[Weaver et al. (2019)]{Weaver et al. (2019)} Weaver, Z. R., Balonek, T. J., et al., 2019, ApJ, 875, 15
 \bibitem[Zacharias et al. (2016)]{Zacharias et al. (2016)} Zacharias, M., B\"ottcher, M., Chakraborty, N., et al. 2016, arXiv:1611.02098
 \bibitem[Zhang et al. (1999)]{Zhang et al. (1999)} Zhang, Y. H., Celotti, A., Treves, A., Chiappetti, L., et al. 1999, ApJ, 527, 719
 \end{thebibliography}

\end{document}